# How to Fix Silver for Plasmonics

*Björn Ewald[1,‡]\*, Leo Siebigs[2,‡], Cheng Zhang[2,‡]\*, Jonas Graf[1], Achyut Tiwari[3], Maximilian Rödel[1], Sebastian Hammer[1], Vladimir Stepanenko[4], Frank Würthner[4], Bruno Gompf[3], Bert Hecht[2]\*, Jens Pflaum[1,5]\**

[1] Experimental Physics 6, University of Würzburg, Am Hubland, 97074 Würzburg, Germany

[2] Experimental Physics 5, University of Würzburg, Am Hubland, 97074 Würzburg, Germany

[3] 1. Physikalisches Institut, Universität Stuttgart, Pfaffenwaldring 57, 70569 Stuttgart, Germany

[4] Institut für Organische Chemie and Center for Nanosystems Chemistry, Universität Würzburg, Am Hubland, 97074 Würzburg, Germany

[5] Center for Applied Energy Research e.V. (CAE Bayern), Magdalene-Schoch-Straße 3, 97074 Würzburg, Germany

[‡] These authors contributed equally

\*Corresponding authors: bjoern.ewald@uni-wuerzburg.de, cheng.zhang@uni-wuerzburg.de, bert.hecht@uni-wuerzburg.de, jens.pflaum@uni-wuerzburg.de

**Abstract**

Silver (Ag) is considered an ideal material for plasmonic applications in the visible wavelength regime due to its superior optical properties, but its use is limited by the poor chemical stability and structural quality of thermally evaporated thin films and resulting nanostructures. In this study, we present a simple approach to enhance the structural and optical quality as well as the chemical stability of Ag thin films by alloying with gold (Au) through thermal co-evaporation. We investigate $Ag_{100-x}Au_x$ thin films with Au contents ranging from 5 to 20 at% analyzing their surface morphology, crystallite structure, optical properties, and chemical stability. Our results show that low Au concentrations significantly reduce the roughness of co-evaporated thin films (down to 0.4 nm RMS), and significantly enhance the resistance to oxidation, while maintaining a defined crystallite growth. Importantly, these improvements are achieved without the need for template stripping, metallic wetting layers, or epitaxial substrates, enabling direct deposition on glass. Among the compositions studied, $Ag_{95}Au_5$ thin films exhibit the highest chemical stability, lowest optical losses in the visible spectral range, and excellent plasmonic properties even outcompeting pure Ag. As a proof-of-concept, we fabricate high-quality $Ag_{95}Au_5$ optical antennas that exhibit long-term durability under ambient conditions. Our approach provides a practical solution to overcome the limitations of Ag for plasmonic device applications.

**Introduction**

Silver (Ag) is considered an ideal plasmonic material at visible wavelengths due to its superior optical properties, including higher reflectivity and lower ohmic losses compared to gold (Au). In particular, Ag is not affected by interband transitions in the blue and green regions of the visible spectrum, as is the case for Au.[1-4] However, its application in polycrystalline thin films is limited by structural imperfections, such as a considerable surface roughness, and rapid degradation under ambient conditions. As a result, Au remains the most widely used optical as well as plasmonic material under ambient experimental conditions.[1, 2, 5]

In terms of process scalability and fabrication ease, physical vapor deposition of metals via thermal evaporation is the preferred method for producing plasmonic thin films, nanostructures and devices.[6, 7] Nucleation, island growth, and coalescence influence the resulting crystallite structure and surface morphology. Moreover, high supersaturation levels and the presence of impurities might lead to intrinsic and extrinsic defects within the deposited structures.[8] Achieving low surface roughness and well-defined crystallite growth is particularly challenging for Ag, due to fast surface atom diffusion and its tendency to react with oxygen and water during deposition.[6, 8] Metal thin films with low surface roughness, oriented crystallite growth, large grain sizes, and low defect densities not only minimize optical losses[9], but also enable the fabrication of well-defined nanostructures using electron-beam lithography (EBL)[7, 10] and focused-ion-beam (FIB) lithography[11, 12]. Several strategies have been developed for depositing high-quality plasmonic Ag thin films. Epitaxial growth of ultrasmooth "single crystalline" films requires substrates with atomic lattice matching, which are often incompatible with optical device integration.[13-17] High-quality polycrystalline thin films have been achieved through elevated deposition rates[4], substrate cooling during deposition[18], and the use of metallic wetting layers[19, 20]. Still, ultrasmooth plasmonic Ag thin films are most often obtained by post-processing via template-stripping.[21-23]

The use of Ag in plasmonic (device) applications is limited not only by the challenges in controlling surface and crystallite morphology, but also by its poor chemical stability. Degradation through (electro)chemical reactions with oxygen, sulfur, moisture and solvents compromises the stability of (nano-)structures and complicates thin-film handling and, thus, hampers technological application.[24-26] Noble metal alloys, particularly $Ag_{100-x}Au_x$ alloys, are of interest both for tuning the material`s dielectric function for practical applications[27-30], and for enhancing the chemical stability of plasmonic structures and thin films.[31, 32] Au as alloying material offers complete miscibility with Ag and high-quality thin film growth.

In this study, we demonstrate that alloying Ag thin films with small proportions of Au via thermal co-evaporation is an effective, straightforward approach to enhance the chemical stability while simultaneously improving surface roughness and crystallite quality. While Gong et al.[28] have demonstrated that high Au contents significantly alter the band structure of $Ag_{100-x}Au_x$ thin films, our objective is different. We aim to preserve the excellent optical properties of Ag in the visible spectral range, which are essential for plasmonic applications, while simultaneously leveraging the improved surface morphology and chemical stability introduced by low-level Au alloying. Notably, our fabrication method does not require template-stripping, metallic wetting layers, epitaxial substrates or any other efforts, thus enabling the direct preparation of application-ready thin films on glass. We provide a comprehensive analysis of the alloying effects, in an Au concentration range from x = 5 to 20 at%, by highlighting the interrelations between structural and optical characteristics of $Ag_{100-x}Au_x$ thin films. Among these, $Ag_{95}Au_5$ films exhibit significantly improved chemical stability compared to pure Ag, along with remarkably low surface roughness (down to 0.4 nm), well-defined crystallite growth, and enhanced optical and plasmonic performance in the visible spectral range. As a proof-of-concept, we demonstrate the functionality of $Ag_{95}Au_5$ structures as plasmonic nanoantennas with excellent resonant properties and long-term ambient stability.

**Results and Discussion**

**Morphology and Crystallite Structure.** Alloying Ag thin films with low concentrations of Au enables significant tuning of surface morphology and crystallite structure and, thus, thin film quality, providing a thin film platform excellently suited for plasmonic applications.

Before alloying, we optimized the deposition parameters for pure Ag thin films (film thickness: 40 to 50 nm), which is described in SI 1 (Figures S1 to S3). To ensure comparability, all $Ag_{100-x}Au_x$ thin films discussed below were deposited under identical conditions: a base pressure below $5 \cdot 10^{-7}$ mbar, a deposition rate of 3 to 4 nm·s$^{-1}$, and a substrate temperature of about 140 K. Prior to deposition, the glass substrates were functionalized with (3-Aminopropyl)triethoxysilane (APTES) serving as wetting layer.

The $Ag_{100-x}Au_x$ alloy composition is achieved via co-evaporation of Ag and Au from separate boat sources by adjusting the respective ratio of the individual deposition rates. The actual alloy composition and the homogenous distribution of Au within the thin films are subsequently verified by energy-dispersive X-ray (EDX) spectroscopy (see SI2, Figure S4). All thin films are labeled according to their actual composition as determined by EDX ($\pm$ 1at%). To analyze the surface and bulk morphology, we employ tapping mode atomic force microscopy (AFM), scanning electron microscopy (SEM) with in-lens detection, X-ray reflectivity (XRR) and X-ray diffraction (XRD).

Representative tapping mode AFM images (area: 10 x 10 μm$^2$) of Ag, $Ag_{95}Au_5$, $Ag_{90}Au_{10}$, $Ag_{80}Au_{20}$ and Au thin films are displayed in Figure 1a. The images clearly demonstrate that even small amounts of Au have a pronounced impact on the thin film surface morphology.

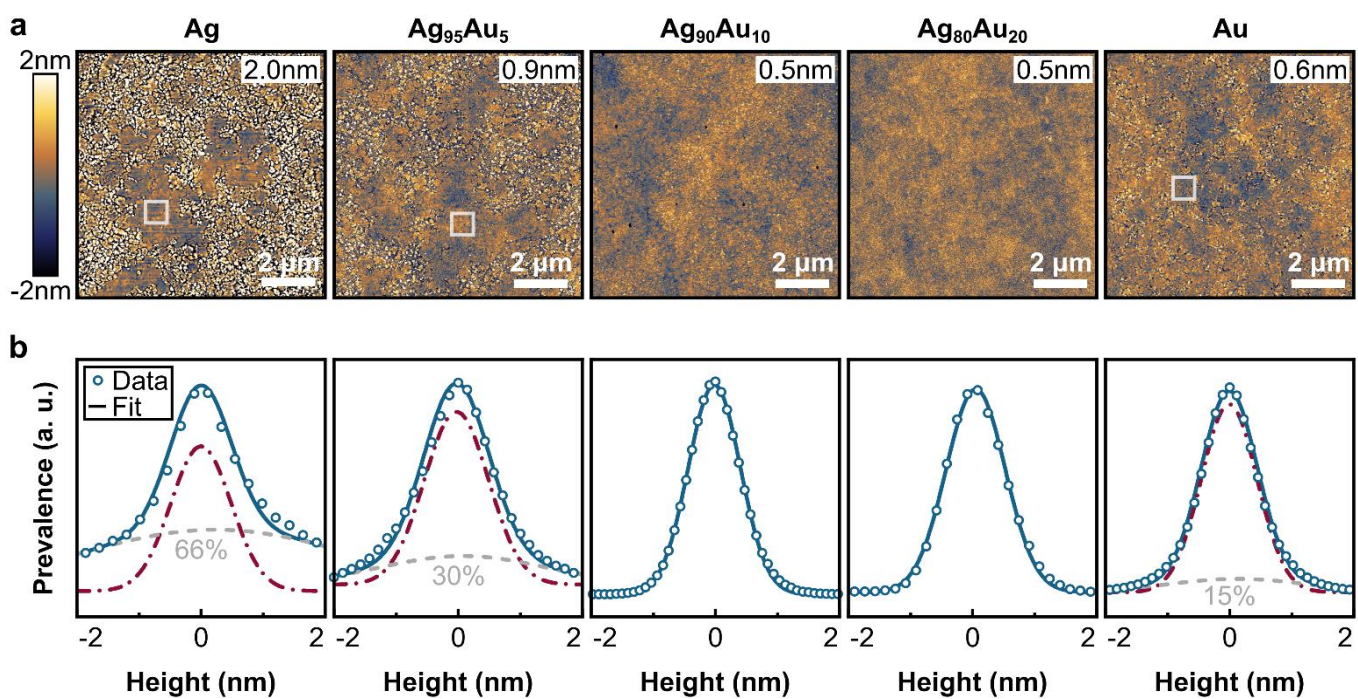

**Figure 1.** Surface morphology of $Ag_{100-x}Au_x$ thin films. Thin film compositions (in at%) are determined by quantitative EDX spectroscopy. a) Tapping mode AFM images of representative 10 × 10 μm² areas. To highlight differences in surface morphology the color scale is normalized from -2 to 2 nm relative to the median height. Respective $Rq$ values of the 10 × 10 μm² areas are shown in the top right. White squares indicate 1 x 1 μm² regions exhibiting ultrasmooth morphology, which are analyzed separately. The surface properties are summarized in Table S1. b) Height distribution functions of the 10 × 10 μm² AFM images in a). For Ag, $Ag_{95}Au_5$ and Au the data were modeled with the sum (blue solid line) of a narrow (dash-dotted black lines, $\sigma$ = 0.5 nm) and a broad (dashed gray lines, $\sigma$ = 2.2 nm, 1.3 nm and 1.1 nm, respectively) Gaussian. Upon alloying, the contribution of the broad distribution decreases from 66 % (Ag) to 30 % ($Ag_{95}Au_5$). In the case of $Ag_{90}Au_{10}$ and $Ag_{80}Au_{20}$ the height distributions are well described by one narrow (blue solid lines, $\sigma$ = 0.4 nm and 0.5 nm, respectively) Gaussian.

The root mean square roughness ($Rq$) is a well-established metric for assessing the surface morphology of thin films. $Rq$ values obtained from the 10 × 10 μm² AFM images are listed shown in Figure 1a and listed in Table S1. Alloying Ag with a small proportion of Au significantly reduces $Rq$ from 2.0 nm (pure Ag) to 0.9 nm ($Ag_{95}Au_5$). With higher Au content, the roughness of $Ag_{90}Au_{10}$ and $Ag_{80}Au_{20}$ films further decreases to 0.5 nm, approaching $Rq$ values observed for pure Au films.

To further investigate the growth mode, height distributions extracted from AFM scans were modeled using Gaussians (see Figure 1b). The standard deviation $\sigma$ of these distributions serves as a measure for surface roughness, analogous to $Rq$. Notably, a single distribution did not yield a satisfactory fit for Ag, $Ag_{95}Au_5$ and Au thin films. The ultrasmooth regions with extended grains, clearly visible in the AFM images (see Figure 1a) contribute to a narrow distribution ($\sigma = 0.5$ nm). A broader distribution with $\sigma = 2.2$ nm (Ag), 1.3 nm ($Ag_{95}Au_5$) and 1.1 nm (Au) is attributed to rougher areas characterized by wide variations in grain size and orientation. This bimodal growth mode is most pronounced for pure Ag, where the broad distribution accounts for 66 % of the $10 \times 10$ μm$^2$ surface area. However, alloying with just 5 at% of Au already significantly narrows the broad distribution, and reduces its area share to 30 %. The surface morphology of $Ag_{95}Au_5$ thus becomes more similar to that of pure Au, which shows an ultrasmooth surface morphology with only 15 % of the surface area corresponding to the broad distribution. Flat regions, separately analyzed by selecting $1 \times 1$ μm$^2$ areas (indicated in Figure 1a), exhibit exceptionally low $Rq$ values of 0.3 nm (Ag), 0.4 nm ($Ag_{95}Au_5$) and 0.5 nm (Au) (see Table S1) which as expected closely align with the standard deviations of the narrow Gaussians used to model the height distributions. Towards slightly more balanced concentrations of $Ag_{90}Au_{10}$ and $Ag_{80}Au_{20}$ the overall thin film roughness is described by just a single, narrow Gaussian distribution with $\sigma = 0.4$ nm and 0.5 nm, respectively.

We further analyzed the grain texture by height-height correlation analysis (SI2, Table S1, Figure S5) and in-lens SEM micrographs (SI2, Figure S6), both confirming the strong impact of Au alloying. Alloying with 5 to 10 at% Au leads to the formation of laterally extended grains, reaching several micrometers in dimension. In contrast, the $Ag_{80}Au_{20}$ thin film stands out, due to its fine-grained morphology.

To obtain integral information on both surface as well as interface roughness over the entire 22 x 22 mm$^2$ sample area, we performed XRR measurements. XRR data for the complete series of Au concentrations studied are displayed in Figure S7. The results confirm our previous conclusions based on AFM on the local surface morphology and further highlight the exceptional interface and surface properties of the $Ag_{95}Au_5$ thin film with *Rq* reduced by a factor of 1/3 compared to Ag. This demonstrates that alloying Ag films with 5 at% Au improves the surface morphology across the entire thin film area. Further details on modeling parameters are given in SI 2.

Considering surface roughness as a key quality factor, our alloyed Ag thin films are competitive to Ag films fabricated by epitaxial growth[13, 14, 17] or template-stripping.[6, 9, 21] Notably, a small Au content of 5 at% leads to an improved surface morphology characterized by low roughness and laterally extended flat grains.

Alloying affects not only the surface morphology but likewise the crystalline structure of the thin films, which we have investigated using out-of-plane XRD measurements (see Figure 2).

Via XRD, we ascertain the fundamental crystallographic properties based on Bragg´s law:

$$q_z = n \frac{2\pi}{d_{hkl}} \quad (1)$$

Here $q_z$ is the out-of-plane momentum transfer, *n* is the diffraction order, and $d_{hkl}$ is the spacing of the respective lattice planes with Miller indices (*hkl*).

Although both Ag and Au crystallize in a face-centered cubic (fcc) structure with only slightly different lattice constants[33], alloying has a strong impact on the formation and spatial extent of crystalline regions within the $Ag_{100-x}Au_x$ films as evidenced by XRD patterns (see Figure 2a). We attribute this to the non-equilibrium conditions inherent to thin film growth during thermal evaporation. The thin films exhibit a textured growth, with well-defined crystallites oriented perpendicular to the substrate plane, evidenced by the distinct peaks attributed to the (111) and (222) Bragg reflections. The peak area is indicative for the crystalline volume fraction present within the films. According to the XRD patterns, $Ag_{95}Au_5$ exhibits the most pronounced crystallinity among all alloyed films, second only to pure Au, which correlates well with the flat extended crystallites observed by AFM and SEM. Alloying with 5 at% Au therefore is effective not only in improving the surface morphology but also in sustaining the long-range order within the thin film. Conversely, higher Au contents appear to impede formation of crystalline domains, as reflected by the diminished peak intensity for $Ag_{90}Au_{10}$ and its nearly complete suppression for $Ag_{80}Au_{20}$. This might be related to strain within the structure induced by alloying. This finding is consistent with the fine-grained texture apparent in the AFM and SEM images for $Ag_{80}Au_{20}$.

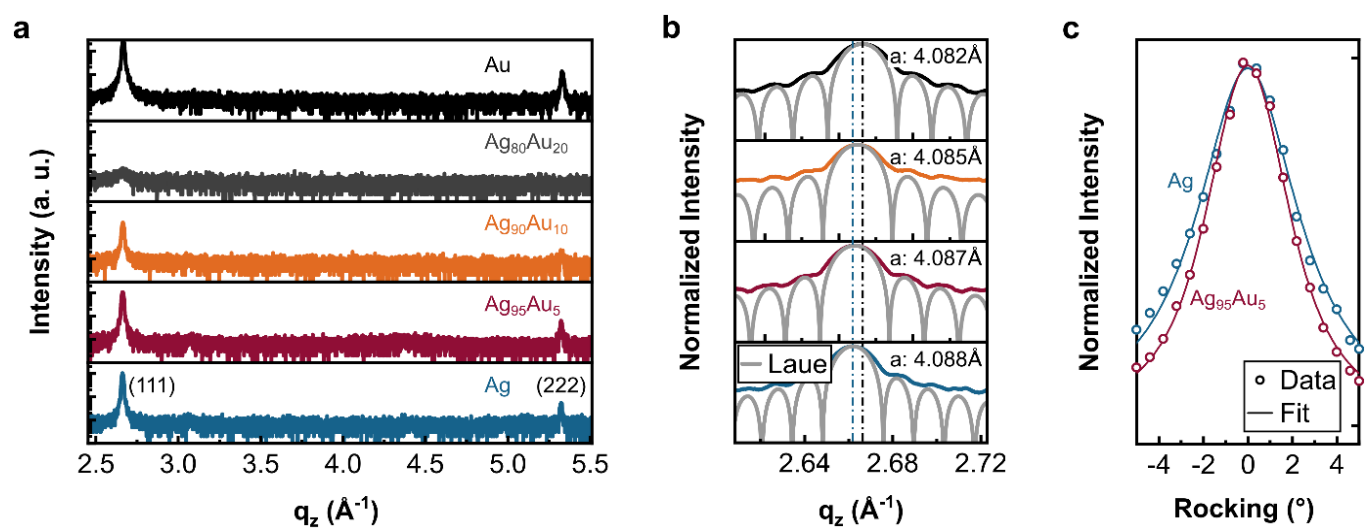

**Figure 2.** Structural characterization of $Ag_{100-x}Au_x$ thin films via out-of-plane X-ray diffraction (XRD). a) XRD patterns ($\theta$-$2\theta$ scan) of the pure and alloyed thin films. The distinct Bragg peaks can be assigned to the (111) and (222) reflections, confirming a crystalline long-range ordering perpendicular to the substrate plane. b) Semilogarithmic normalized (111) Bragg reflection intensity, obtained from high-resolution scans for Ag, $Ag_{95}Au_5$, $Ag_{90}Au_{10}$ and Au thin films. The presence of Laue oscillations indicates high crystallite quality and low defect density. The simulation (grey lines) matches the periodicity and intensity distribution observed in the experiment, under the assumption of coherently scattering domain sizes of 40 nm (170 lattice planes), i.e. close to the respective film thicknesses, as determined by XRR. The fcc-lattice constant $a$ derived from the position of the (111) Bragg reflection is shifting from 4.088 Å (pure Ag) to 4.082 Å (pure Au) with increasing Au content. c) Rocking curves on the (111) Bragg reflection for Ag and $Ag_{95}Au_5$ thin films. Voigt functions were fitted to the data and the full-width-at half maximum decreases from 5.6° for Ag to 4.5° for $Ag_{95}Au_5$. This suggests a lower mosaicity, i.e. better defined out-of-plane alignment of the crystalline regions, for the $Ag_{95}Au_5$ thin film.

To gain deeper insight and establish a quantitative relationship between concentration and crystalline properties, we performed a detailed analysis of the (111) Bragg reflections (see Figure 2b). The data reveal an exponential shift in the lattice parameter with increasing Au content. Using Equation (1), we find that the fcc-lattice constant $a = d_{111} \cdot 3^{1/2}$ decreases from 4.088 Å (pure Ag) to 4.082 Å (pure Au). The occurrence of Laue oscillations on both sides of the primary peak, is an indication for the long range ordering of lattice planes and, thus, for the low defect density within the crystallites. The intensity $I(q_z)$ of these subsidiary maxima in the vicinity of the main Bragg reflection is characterized by the Laue interference function:[34]

$$I(q_z) \propto \frac{sin\left(\frac{N}{2}q_z d_{hkl}\right)^2}{sin\left(\frac{1}{2}q_z d_{hkl}\right)^2} \qquad (2)$$

Assuming a coherently diffracting out-of-plane volume of 40 nm, which is close to the thin film thicknesses determined by XRR (see Table S2) and corresponds to about 170 lattice planes, the Laue interference function reproduces the experimental intensity distributions quite well (see Figure 2b). This suggests that crystalline domains forming along the out-of-plane direction extend across the entire film thickness and are hardly affected by structural inhomogeneities.

Finally, we analyzed the crystallite tilting with respect to the surface normal, i.e. the mosaicity, by rocking curve measurements on the (111) Bragg reflection (see Figure 2c and Figure S8). A Voigt function was employed to model the experimental data, yielding a full width at half maximum of 5.6° for pure Ag and of 4.5° for $Ag_{95}Au_5$. This shows that the film's mosaicity is reduced by alloying Ag with a small amount of Au, resulting in a more oriented crystallite growth.

Consequently, based on the presented structural data, alloying Ag thin films with as little as 5 at% Au proves to be an effective strategy for simultaneously enhancing the surface morphology and optimizing the crystallite structure. In the following, we focus on Ag, $Ag_{95}Au_5$, $Ag_{80}Au_{20}$ and Au thin films to analyze the optical and plasmonic properties.

**Optical and Plasmonic Properties.** With respect to plasmonic applications the optical performance of our thin films (see Figure 3) is of central importance. To identify the effect of chemical composition on the interband transitions, we recorded solid-state transmission and reflection spectra of the $Ag_{100-x}Au_x$ alloy films (see Figure S9), with the corresponding normalized absorption spectra shown in Figure 3a. As the Au content increases, the onset of absorption associated with Au specific interband transitions between the d- and sp-band shifts to lower energies. Importantly, the absorption spectrum of the $Ag_{95}Au_5$ layer reveals that the visible range below 3.10 eV (400 nm) is unaffected for alloying levels up to 5 at%, preserving the desirable optical characteristics of Ag.

The interaction of metals with electromagnetic radiation is governed by the complex dielectric function $\varepsilon = \varepsilon_1 + i\varepsilon_2$. The real part ($\varepsilon_1$) of the dielectric function defines the response of the electrons to the electric field, while the imaginary part ($\varepsilon_2$) accounts for optical losses arising from damping of the electrons' dynamics. $\varepsilon_2$ is highly sensitive to the crystallinity and crystallite structure of metal layers.[9] We applied spectroscopic ellipsometry to determine the dielectric function of $Ag_{100-x}Au_x$ thin films in a broad spectral range (210 nm to 1600 nm). The extracted real and imaginary parts of the dielectric functions in the visible spectral range are displayed in Figure 3b and c, the complete spectra are provided in Figure S10.

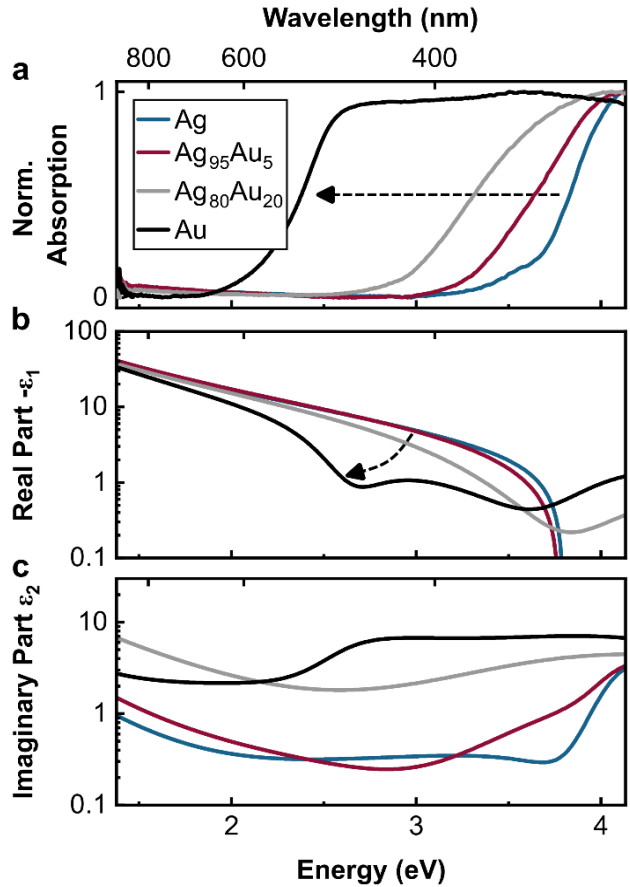

**Figure 3.** Optical properties of $Ag_{100-x}Au_x$ thin films. a) Normalized absorption spectra showing the red-shift of interband transitions with increasing Au content. b) Negative real part and c) imaginary part of the dielectric functions in semilogarithmic presentation. The real and imaginary part of the complex dielectric function $\varepsilon = \varepsilon_1 + i\varepsilon_2$ were obtained using spectroscopic ellipsometry. The imaginary part of $Ag_{95}Au_5$ resembles that of pure Ag. In contrast, the low degree of crystallinity, and the fine-grained texture of $Ag_{80}Au_{20}$ contribute to increased optical losses (larger $\varepsilon_2$).

Interestingly, Ag and $Ag_{95}Au_5$ feature a similar response in the visible spectral range with almost identical imaginary parts below 1 and an even more negative real part in the case of $Ag_{95}Au_5$. The excellent optical properties of Ag are thus maintained and even improved through minor Au alloying, which is most likely due to lower roughness, more oriented growth and lower defect density. As summarized in Table S3 the optical properties of our Ag and $Ag_{95}Au_5$ thin films are comparable to leading reference systems reported in the literature[4, 6, 9], including those fabricated via template-stripping[5, 8]. In contrast the high $\varepsilon_2$ values observed for the $Ag_{80}Au_{20}$ thin film reflect its disturbed long-range order and its fine-grained texture leading to increased damping.

To access the plasmonic characteristics, we utilize a Kretschman configuration[35-37] for probing the surface plasmon polariton (SPP) dispersions of $Ag_{100-x}Au_x$ thin films on the glass substrate. P-polarized white light from a halogen lamp is passed through a hemispherical lens coupled to the glass substrate of the thin metal sample via refractive index matching immersion oil (see a simplified sketch in Figure S11). SPP excitation occurs at the metal-air interface via the evanescent field of light reflected at the glass-metal interface. The wavevector ($k$)-dependent SPP dispersions of the thin films are displayed in Figure 4a, and the complementary angle-dependent data are shown in Figure S12. The curvature and width of the SPP dispersions are found to be significantly influenced by the thin film composition. The widths of the SPP dispersions are compared by means of $k$-dependent cross-sections at 2.07 eV (600 nm) and 2.48 eV (500 nm) (see Figure 4b). The $Ag_{95}Au_5$ thin film exhibits the narrowest peak at both energies, indicating the lowest optical losses and thus, the highest plasmon propagation length. This behavior is consistent with its superior dielectric response and exceptionally smooth surface morphology. Conversely the high optical loss and interband transitions lead to a widening of the SPP dispersion in the case of $Ag_{80}Au_{20}$ and Au. Additionally, the composition-dependent SPP dispersions reflects changes in the high-frequency dielectric constant $\varepsilon_\infty$, which increases linearly with Au content (see Figure 4c). The values of $\varepsilon_\infty$ were obtained from fits to the experimental dispersion relations (see Figure S13) and represent the influence of interband transitions on the plasmonic response.[9, 28]

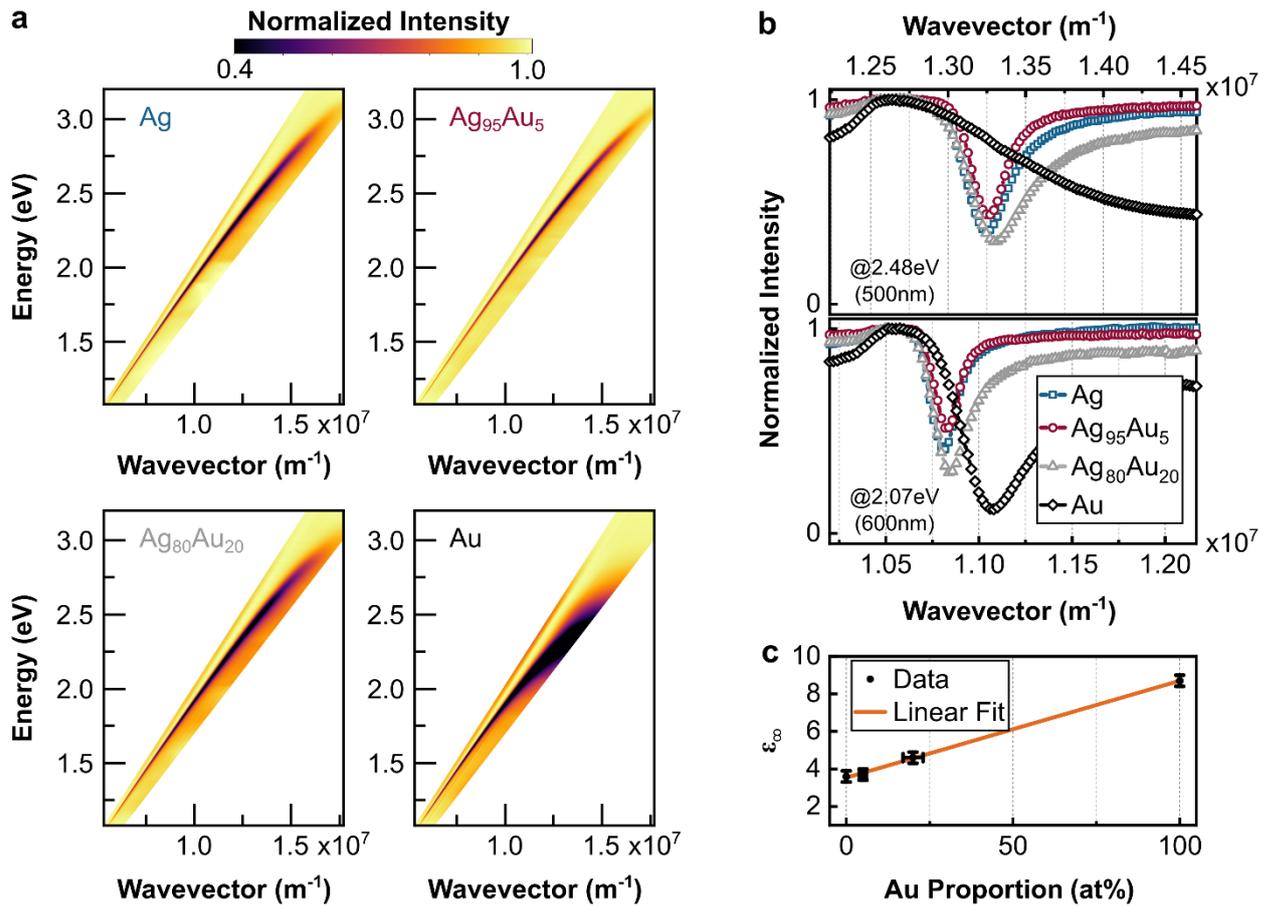

**Figure 4.** Surface plasmon polariton (SPP) dispersions of $Ag_{100-x}Au_x$ thin films. a) SPP dispersions measured in Kretschmann configuration. The color scale represents the normalized reflected intensity. An intensity minimum is caused by SPP excitation at the metal-air interface. b) Cross-sections at 2.07 eV (600 nm) and 2.48 eV (500 nm). The $Ag_{95}Au_5$ thin film exhibits the narrowest reflection peak and thus the smallest optical losses and highest plasmon propagation lengths at both energies. This can be directly attributed to the remarkable optical properties and the ultralow surface roughness. c) High-frequency dielectric constant $\varepsilon_\infty$ as function of the Au content, derived from fits to the SPP dispersion relations. $\varepsilon_\infty$ increases linearly with rising Au proportion, reflecting the influence of interband transitions.

**Oxidative Stability.** To implement Ag as plasmonic material, in physiological or technological environments the long-term stability and inertness against oxidation are of pivotal importance. However, pure Ag is susceptible to (electro-)chemical reactions, leading to surface and bulk degradation of thin films and nanostructures and hence of their optical properties. Thus, robust Au remains the most widely applied plasmonic material despite its inferior optical properties. The objective of alloying Ag with Au is to achieve a significantly enhanced oxidative stability, while keeping the Au content below 10 at%. Thereby we preserve the advantageous morphological and plasmonic properties, as discussed above. To evaluate oxidative degradation, we have exposed Ag, Ag$_{95}$Au$_5$, Ag$_{80}$Au$_{20}$ and Au thin films to an oxidative environment by means of immersion in a 0.2 % (w/w) aqueous H$_2$O$_2$ solution for 15 minutes. The SEM micrographs and XRR data for the full series are discussed in SI 4 (Figure S14). SEM micrographs of Ag and Ag$_{95}$Au$_5$ before and after the degradation test are presented in Figure 5.

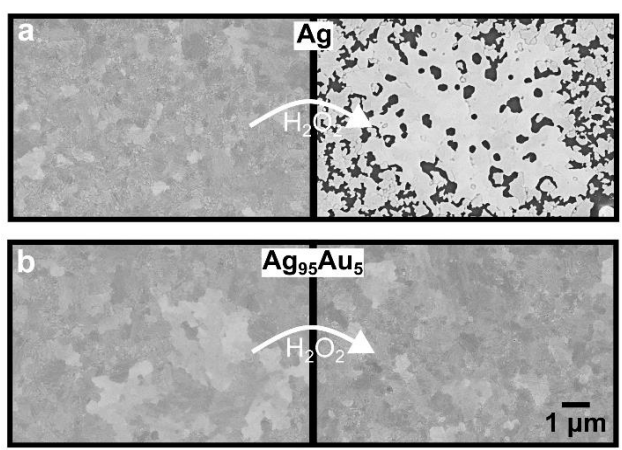

**Figure 5.** Oxidative stability of Ag (a)) and Ag$_{95}$Au$_5$ (b)) thin films. The thin films were immersed for 15 minutes into 0.2 % (w/w) aqueous H$_2$O$_2$ solution. Representative in-lens SEM micrographs (acceleration voltage 5 kV) are shown before (left) and after immersion (right). A markedly enhanced stability is observed for the Ag$_{95}$Au$_5$ thin film compared to pure Ag. The scale bar is 1µm.

As demonstrated by the SEM micrographs (Figure 5a), the pure Ag film is strongly affected by oxidation, which is apparent by the removal of significant film parts from the substrate and the substantial alteration of the grain texture of the remaining film. These changes are corroborated by the XRR measurements (Figure S14 b), where a shift of the critical angle of total reflection to lower $q_z$-values indicates a reduction in the film´s integral electron density ($q_{crit} \sim (\rho_{el})^{1/2}$ due to material loss and incorporation of oxygen atoms (2Ag + H$_2$O$_2$ → Ag$_2$O + H$_2$O). Impressively though, the presence of small amounts of Au in Ag$_{95}$Au$_5$ thin films significantly enhances the stability against oxidation. In the SEM micrographs (Figure 5b), we neither see indication of material removal nor of grain texture modification. Furthermore, the edge of total reflection in the XRR measurements (Figure S14d) remains constant, indicating that no significant amounts of oxygen are incorporated and no substantial amounts of Ag are dissolved or oxidized. The remarkable oxidative stability of $Ag_{95}Au_5$ complements the excellent structural, optical and plasmonic properties. In conclusion we demonstrate the operation and ambient stability of optical nanoantennas based on $Ag_{95}Au_5$.

**Plasmonic Nanoantennas.** Finally, we demonstrate the fabrication of low-loss and environmentally stable plasmonic nanoantennas based on $Ag_{95}Au_5$. The structures were produced using high-precision nanofabrication techniques, including electron-beam lithography (EBL) and helium-ion milling (HIM).

Figure 6a presents a series of plasmonic nanoantennnas featuring length-dependent dipolar resonances fabricated using a HIM protocol,[38] with precise dimension control implementing 30 nm length increments (see SEM images in the top panel). The corresponding white-light scattering (WLS) spectra (Figure 6a, bottom) reveal dipole resonances across the visible spectrum (500 to 800 nm), confirming the length tunability of these nanoantennas fabricated from the alloy. Simulated spectra show good agreement with experimental data, with minor deviations attributed to fabrication tolerances of ± 5 nm in all directions.

To assess long-term ambient stability, we conducted WLS measurements on EBL-fabricated nanoantennas immediately after fabrication and after one month of storage under ambient conditions. Figure 6b compares resonance shifts and signal retention of at least 13 antennas per thin film composition (Ag, $Ag_{95}Au_5$, Au). Stability was quantified via a yield metric defined as the proportion of antennas maintaining at least 20 % of their initial scattering intensity and exhibiting a resonance wavelength shift < 75 nm after one month in ambient conditions. $Ag_{95}Au_5$ antennas exhibit a yield of 78 %, significantly outperforming pure Ag antennas, which show only 23 % yield under identical conditions. This improvement directly reflects the enhanced oxidative stability observed for the corresponding thin films (Figure 5), preserving the optical response. Au antennas, known for their superior chemical inertness, achieved a slightly higher yield of 92 %. These findings establish $Ag_{95}Au_5$ as a promising material for high-performance plasmonic nanostructures, offering ambient stability, low optical losses and ultrasmooth surfaces.

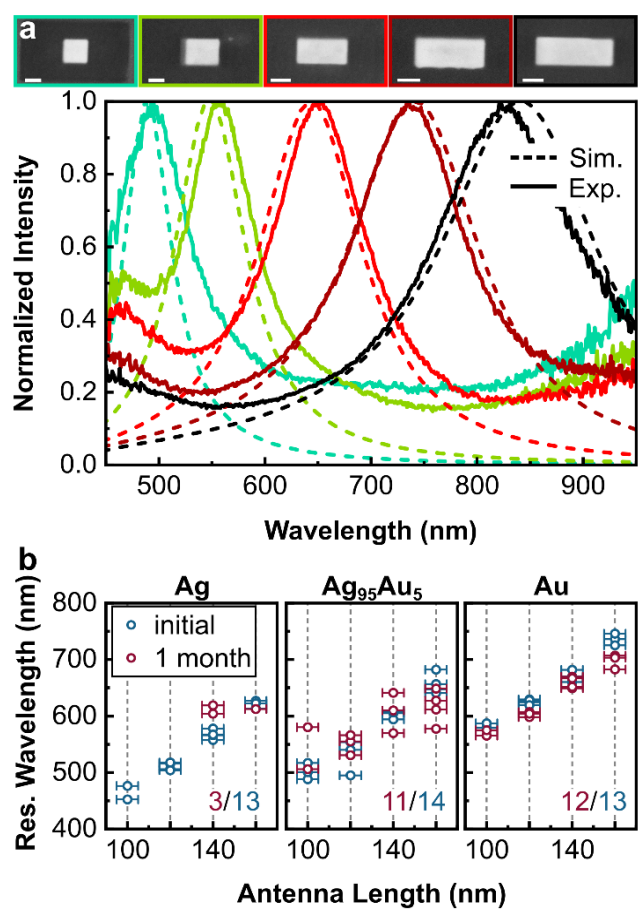

**Figure 6.** Optical nanoantennas based on $Ag_{95}Au_5$. a) White light scattering (WLS) spectra of FIB-fabricated nanoantennas, with corresponding SEM images shown above (scale bar: 50 nm). The Finite-Difference Time-Domain (FDTD) simulations based on the experimentally derived dielectric function of $Ag_{95}Au_5$ (dashed lines) are in good agreement with the measured spectra. Minor deviations in resonance wavelength are attributed to structural uncertainties of ± 5 nm in all directions, b) Durability test of EBL-fabricated Ag, $Ag_{95}Au_5$ and Au nanoantennas. Displayed are the initial resonance wavelengths and after one month under ambient storage. Antennas are considered stable if they exhibit a resonance wavelength shift of < 75 nm and retain > 20 % of their initial scattering intensity. Alloying with 5 at% Au significantly enhances ambient stability, increasing the yield from 23 % (Ag) to 78 % ($Ag_{95}Au_5$), approaching the benchmark stability of Au (92 %).

**Conclusions.** In this work, we have shown that alloying Ag thin films with small amounts of Au via thermal co-evaporation is a simple yet effective strategy to overcome key limitations of Ag in plasmonic applications. By systematically investigating $Ag_{100-x}Au_x$ thin films with Au contents between 5 and 20 at%, we find that alloying at low Au concentrations significantly improves surface morphology, crystallite quality, and chemical stability. This is achieved without the need for epitaxial substrates, template stripping, or metallic wetting layers, allowing direct deposition on glass substrates. Pure Ag thin films exhibit an RMS roughness of 2.0 nm, which is reduced to as low as 0.4 nm in $Ag_{100-x}Au_x$ films. Additionally, the alloyed films demonstrate high resistance to chemical oxidation, approaching that of pure Au. We identify $Ag_{95}Au_5$ as the optimal composition, offering reduced surface roughness, well-defined crystallite growth, and significantly enhanced chemical stability, while maintaining the low optical losses and the broad spectral accessibility characteristic of Ag. SPP dispersion measurements further reveal that $Ag_{95}Au_5$ exhibits even lower plasmonic losses than pure Ag. As a proof-of-concept, $Ag_{95}Au_5$ nanoantennas show excellent resonant properties and long-term durability under ambient conditions. $Ag_{100-x}Au_x$ thin films are thus ideally suited for plasmonic device applications.

**Experimental Section**

**Thin Film Deposition.** Glass substrates (Karl Hecht, 22 × 22 mm$^2$, thickness 170 μm) were thoroughly cleaned by consecutive ultrasonication (15 minutes each) in double distilled water (CarlRoth) with mucasol® detergent (schuelke), in pure double distilled water, in acetone and in isopropanol, and were dried in a nitrogen stream. Afterwards the glass substrates were functionalized with (3-Aminopropyl)triethoxysilane (APTES, purchased from Sigma Aldrich at 99 % purity). The substrates were placed on the bottom of a 50 mL Erlenmeyer flask with a vial filled with 100 μL APTES located beside the substrate. The flask was flushed with a gentle stream of argon gas (5.0) for at least 60 s and sealed with a septum and parafilm to leave an argon atmosphere above the substrate and the APTES vial during film formation. The sealed flask was placed on a heating plate and functionalization was carried out for 12 to 16 hours at 50 °C. The functionalized substrates were cleaned in an argon stream and directly transferred into the metal deposition chamber. The substrates were fixed on a copper holder to ensure thermal contact to the copper cooling block equipped with a feed and drain line for liquid nitrogen. The cooling process was launched at a base pressure of $8\cdot10^{-7}$ mbar and the depositions were carried out at a base pressure below $5\cdot10^{-7}$ mbar. The deposition was started after 10 minutes of thermal equilibration at 140 K. The deposition rate and thickness were controlled with a quartz crystal microbalance (Kurt J. Lesker, FTM2400-H2R, 6 MHz). Ag was purchased at a purity of 99.99 % (Prof. Feierabend GmbH) and evaporated from self-made tungsten boat sources by resistive heating. Au was purchased at a purity of 99.99 % and evaporated from commercial molybdenum boat sources (Kurt J. Lesker) by resistive heating. The deposition rate was stabilized at 3 to 4 nm·s$^{-1}$ and at least 20 nm of material were evaporated against the shutter prior to actual deposition. For the fabrication of $Ag_{100-x}Au_x$ thin films the alloy ratio was adjusted by co-evaporation from the two boat sources. Once the deposition rate of Au was stabilized the deposition rate of Ag was set to reach the respective alloy composition with an overall deposition rate of 3 to 4 nm·s$^{-1}$. The thin films were deposited at thicknesses of 40 to 50 nm.

**Atomic Force Microscopy.** Atomic force microscopy (AFM) height images were recorded in tapping mode with a Veeco Dimension Icon with standard silicon AFM probes (NanoWorld, NCHR) of 320 kHz resonance frequency and 42 N·m$^{-1}$ force constant. The majority of scans was performed on 10 x 10 μm$^2$ areas with 512 lines per image. Data analysis was carried out with the open software source Gwyddion.[39]

**X-Ray Reflectivity and X-Ray Diffraction.** X-Ray reflectivity (XRR) and X-Ray diffraction (XRD) measurements of the thin films were carried out with a GE Inspection Technologies XRD 3003 TT diffractometer equipped with a monochromatic CuK$_{\alpha 1}$ source of $\lambda$ = 1.5406 Å. The system operates in Bragg-Brentano geometry with a horizontally fixed sample. The source and detector goniometer arms rotate on a Rowland-circle in $\theta$-$\theta$ geometry. XRR measurements were performed at small incident angles up to 5° (2$\theta$) with an angle step size of 0.005° and an integration time of 2 s. XRD scans were performed from 35 to 85° with an angle step size of 0.010° and an integration time of 10 s. High-resolution scans from 37 to 39° were performed with an angle step size of 0.002° and an integration time of 10 s. Rocking curves were measured keeping the $\theta$-2$\theta$ condition of the Bragg (111) reflection fixed while rotating both arms around the sample with an angle step size of 0.010° and an integration time of 5 s. A moving average filter (10 data points) was applied to the (111) Bragg reflections and rocking curves prior to data evaluation.

**Scanning Electron Microscopy.** Scanning electron microscopy (SEM) micrographs of the thin films were measured at a Zeiss SEM Gemini 450 with in-lens detection at an acceleration voltage of 5 kV and a magnification of 10,000.

**Energy-Dispersive X-Ray Spectroscopy.** Energy dispersive X-Ray (EDX) spectroscopy was performed at a Zeiss Ultra Plus field emission scanning electron microscope equipped with a GEMINI e-Beam column. An Oxford X-MAX EDX detector, an acceleration voltage of 10 kV and an aperture size of 120 μm were applied for the quantitative analysis and the qualitative mapping of Ag and Au.

**UV-Vis Spectroscopy.** UV-Vis measurements of the thin films were carried out with a JASCO V-650 spectrophotometer equipped with an integrating sphere JASCO ISV-722. The measurements were performed in a wavelength regime from 300 to 900 nm, with a bandwidth and resolution of 1 nm, under correction for dark and background spectra. Transmission and reflection spectra of the samples were measured by placing the sample in front or behind the integrating sphere, respectively. The glass substrate was measured separately. The absorption spectra of the thin films are then calculated on basis of the sample and substrate spectra.

**SPP Dispersion Measurement in Kretschmann Geometry.** White light from a Thorlabs halogen lamp (SLS201L/M) is passed through a hemispherical N-BK 7 lens and reflected at its plane backside, where the sample is placed. P-polarization is achieved by combining two polarizing beamsplitters CCM1-PBS251/M and CM05-PBS202, respectively. The reflected light is detected by a MAYA2000Pro spectrometer from Ocean Optics with a spectral range from 300 nm to 1100 nm. Index matching between sample substrate and hemispherical lens is ensured by immersion oil (Olympus, IMMOIL F30CC). The SPP dispersion is measured in $\theta$-$2\theta$ configuration. The wavevector $k$ is then given by the following relation to the angle of incidence $\theta$:

$$k = \frac{2\pi n}{\lambda} sin(\theta) \qquad (3)$$

Here $n$ is the refractive index of the hemispherical lens and the immersion oil, $\lambda$ is the wavelength of the light, and $\theta$ is the incident angle as defined by the homebuilt Kretschmann setup.[37]

**Spectroscopic Ellipsometry.** Spectroscopic ellipsometry measurements were conducted using a J.A. Woollam RC2-UI vertical ellipsometer equipped with a dual rotating compensator system. The instrument operated over a spectral range of 210 to 1690 nm. Variable angle measurements were acquired at multiple angles of incidence from 25° to 65° in steps of 5°. Ellipsometry characterizes the change in polarization upon reflection, quantifying the amplitude ratio ($\Psi$) and phase difference ($\Delta$) between p- and s-polarized light. From these parameters, the complex dielectric function ($\varepsilon = \varepsilon_1 + i\varepsilon_2$), representing the real ($\varepsilon_1$) and imaginary ($\varepsilon_2$) components of the dielectric response, was extracted. Data analysis employed a multilayer optical model comprising a glass substrate with an overlying $Ag_{100-x}Au_x$ alloy layer. Optical modeling was based on Fresnel's equations and implemented via regression fitting to match the experimental $\Psi$ and $\Delta$ spectra. All data fitting and analysis were carried out using the CompleteEASE 6 software package (J.A. Woollam Co., Inc.).

**Degradation Tests.** Thin films were immersed (15 minutes) in a 0.2 % (w/w) aqueous $H_2O_2$ solution with ultrasonic agitation to simulate oxidative aging. SEM and XRR measurements were carried out before and after immersion.

**Fabrication of Plasmonic Nanoantennas.** *Electron beam lithography (EBL):* The glass substrate was ultrasonically cleaned in deionized water, acetone, and IPA (10 min each), followed by 10-minute oxygen plasma treatment (250 W, 20 sccm $O_2$). Double PMMA layers (150 nm 600K + 50 nm 950K, Allresist GmbH) were spin-coated and baked at 150°C for 3 min. A 10 nm Au discharging layer was thermally evaporated. Electron-beam patterning was performed using a Zeiss SEM Gemini 450 (30 kV, 35 pA, dose: 1400 $\mu C \cdot cm^{-2}$). A cold development protocol was applied to ensure high resolution pattern: (1) immersion in MIBK:IPA (3:1, 120 s) around -5°C; (2) immersion in IPA 30s (around -5°C) for development termination; (3) $N_2$ flow drying the sample. APTES and the metal thin film were deposited according to the protocol described above. Finally, the sample was placed in acetone for 15 minutes with ultrasonic agitation to complete the lift-off. *Focused-ion-beam milling (FIB):* The freshly evaporated $Ag_{95}Au_5$ thin film was transferred into a FIB microscope. Coarse milling with $Ga^+$ ions (30 kV, 30 pA) was applied for defining the rough nanoantenna geometry, followed by fine nanostructuring using $He^+$ ions (35 kV, 3 pA) to achieve the final nanoantenna dimensions.[38]

**White Light Scattering.** White-light scattering (WLS) spectra measurements were performed using a homebuilt setup.[40] We used a multi-mode fiber coupled from a halogen lamp as an excitation light source. The light beam is then focused into the back focal plane of an oil-immersion objective (Plan-Apochromat, 100x, NA =1.45, Olympus) in order to illuminate the sample with a collimated beam. A beam blocker is inserted in the detection path which solely allows for the collection of scattered light. The scattered light is captured by a spectrometer (Shamrock 303i) equipped with a sensitive EMCCD (iXon A-DU897-DC-BVF, Andor).

**Numerical Simulations.** Three-dimensional finite-difference time-domain (FDTD) simulations were performed using the Maxwell equation solver in the commercial software Ansys Lumerical. A high-resolution mesh (1 nm in all dimensions) was applied to the plasmonic nanoantenna region to ensure accuracy. Perfectly matched layers (PML) were applied at all boundaries to suppress back reflections. To approximate infinite geometry the glass substrate was extended into the PML regions. The total field scattered field (TFSF) with linear polarization along the long antenna axis was used as the source. The antenna is placed above a glass substrate on top of a glass pedestal of $h_p$ = 10 nm height. All antennas are simulated with a thickness $d$ = 40 nm, width 75 nm and different lengths $l$ (85 nm, 100 nm, 130 nm, 160 nm, 190 nm). Rounding of the edges and corners was considered with a radius of $r_c$ = 10 nm.

AUTHOR INFORMATION

Author Contributions

The manuscript was written through contributions of all authors. All authors have given approval to the final version of the manuscript. ‡These authors contributed equally.

Funding Sources

This work was supported by the German Research Foundation (projects HE 5618/8-1 and PF385/12-1) and by the Bavarian State Ministry for Research and the Arts (Collaborative Research Network "Solar Technologies Go Hybrid" (SolTech).

Notes

The authors declare no conflicts of interest.

ACKNOWLEDGMENT

B.E., J.G., M.R., J.P. acknowledge financial support by the Bavarian State Ministry for Science and the Arts within the collaborative research network "Solar Technologies go Hybrid" (SolTech). S.H. acknowledges financial support by the Deutsche Forschungsgemeinschaft (DFG, German Research Foundation) under Germany's Excellence Strategy through the Würzburg-Dresden Cluster of Excellence on Complexity and Topology in Quantum Matter - ct.qmat (EXC2147, project-id 390858490).

# Supporting Information for

# How to Fix Silver for Plasmonics

*Björn Ewald[1,‡]\*, Leo Siebigs[2,‡], Cheng Zhang[2,‡]\*, Jonas Graf[1], Achyut Tiwari[3], Maximilian Rödel[1], Sebastian Hammer[1], Vladimir Stepanenko[4], Frank Würthner[4], Bruno Gompf[3], Bert Hecht[2]\*, Jens Pflaum[1,5]\**

[1] Experimental Physics 6, University of Würzburg, Am Hubland, 97074 Würzburg, Germany

[2] Experimental Physics 5, University of Würzburg, Am Hubland, 97074 Würzburg, Germany

[3] 1. Physikalisches Institut, Universität Stuttgart, Pfaffenwaldring 57, 70569 Stuttgart, Germany

[4] Institut für Organische Chemie, Universität Würzburg, Am Hubland, 97074 Würzburg, Germany

[5] Center for Applied Energy Research e.V. (CAE Bayern), Magdalene-Schoch-Straße 3, 97074 Würzburg, Germany

[‡] These authors contributed equally

\*Corresponding authors: bjoern.ewald@uni-wuerzburg.de, cheng.zhang@uni-wuerzburg.de, bert.hecht@uni-wuerzburg.de, jens.pflaum@uni-wuerzburg.de

## SI. 1: Parameters for Ag Thin Film Deposition

We have optimized the deposition parameters for the fabrication of flat, homogeneous Ag thin films. In order to avoid detrimental reactions of Ag with humidity and oxygen, which interfere crystalline thin film growth by promoting grain boundary formation, all depositions were carried out in high vacuum at a base pressure below $5 \cdot 10^{-7}$ mbar.[1] A monolayer thick (3-aminopropyl)triethoxysilane (APTES) layer on the glass substrates mediates chemisorption of metal (Ag, Au) atoms by the amine group and thus, lowers the tendency for island formation. Furthermore, the adhesion stability of metal thin films and nanostructures is improved.[2,3] The gas-phase functionalization, as outlined in the experimental section, facilitates the deposition of smooth and uniform APTES layers on glass substrates (see AFM image in Figure S1) with an *Rq* roughness of 0.1 nm.

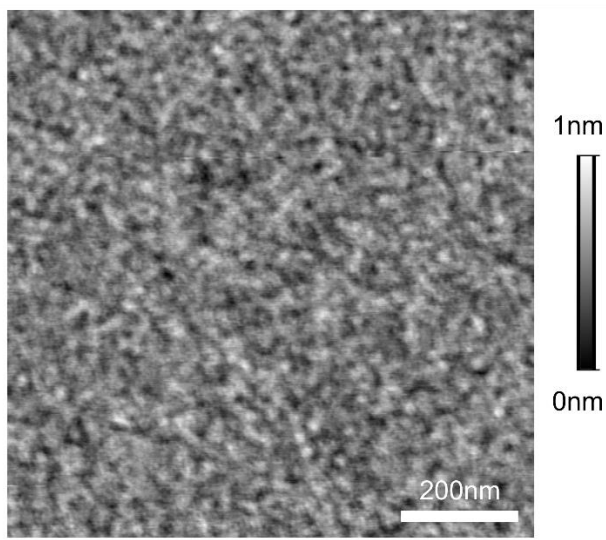

**Figure S1.** Surface morphology of an APTES layer on cover glass measured by tapping mode AFM. The image shows a representative 1 x1 µm² area with *Rq* of 0.1 nm.

As for the results shown in the main section, all thin films were deposited with thicknesses of 40 to 50 nm. The deposition rate exerts a significant influence on the nucleation density and thus, on the propensity for island formation during film growth. Hence, the rate is a key parameter to control the Ag thin film surface morphology. Suppression of surface atom diffusion by elevated deposition rates of 3 nm·s$^{-1}$ leads to a reduction in $Rq$ roughness to 2.9 nm (see Figure S2 a and b), in comparison to lower deposition rates of 0.1 to 2 nm·s$^{-1}$. For elevated deposition rates of 3 to 4 nm·s$^{-1}$ a balance is achieved between stable and controlled evaporation on one hand, and a high surface quality of the resulting thin films on the other. This is a prerequisite for conducting defined Au alloying experiments.

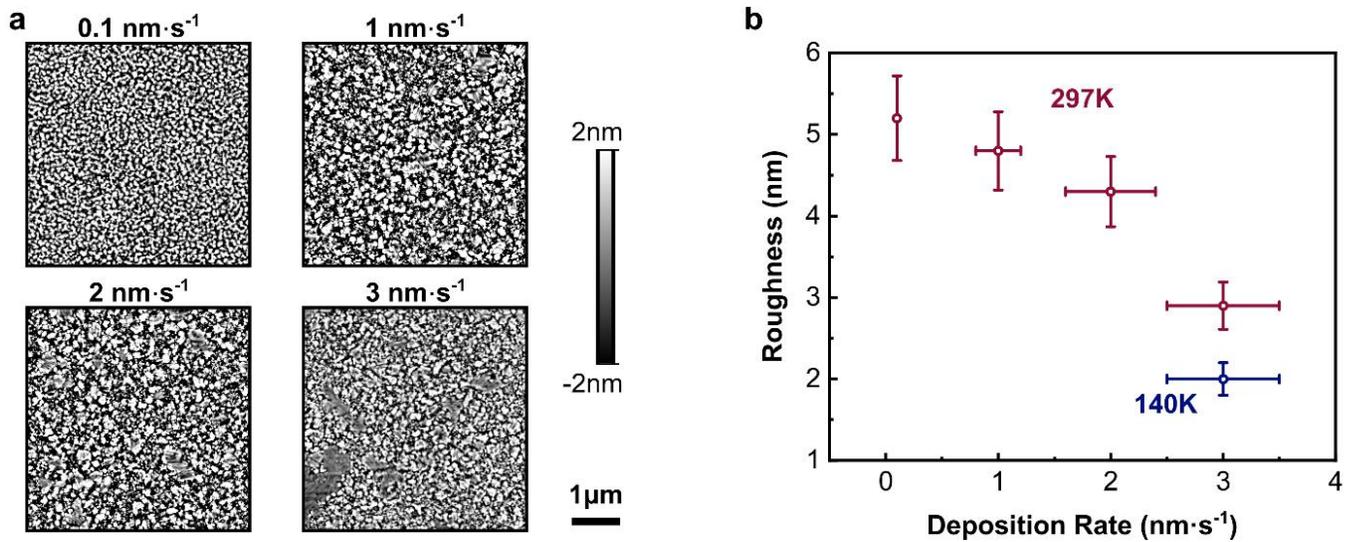

**Figure S2.** Surface morphology of Ag thin films as function of the deposition rate (0.1, 1, 2 and 3 nm·s$^{-1}$). a) Tapping mode AFM images of representative 5 x 5 µm$^2$ sample areas. To emphasize differences in surface morphology, the grey scale is set from -2 to 2 nm relative to the median of the height distribution.. b) Corresponding $Rq$ roughness values plotted against deposition rate. Each error bar reflects variations in surface roughness across three samples and the inherent fluctuations in deposition rate. A significantly improved surface morphology is observed at a higher deposition rate of 3 nm·s$^{-1}$, which still ensures stable and controlled evaporation. As indicated by the blue data point, the surface roughness can be even further reduced by cooling the sample with liquid N$_2$ down to 140 K.

As visible from the AFM images, the formation of flat, extended crystal grains is initiated at elevated deposition rates of 3 nm·s$^{-1}$. Consequently, increasing the nucleation density by limiting surface diffusion of atoms is an effective way to enhance the crystallinity and minimize the surface roughness, provided, that grain boundary formation by excessive moisture or oxygen is avoided. By slowing atom movement via substrate cooling extended island formation can be further suppressed. A comparison of the morphologies of thin films deposited at substrate temperatures of 293 K and 140 K is given in Figure S3. It is evident from the tapping mode AFM images displayed in Figure S3 a that the fraction of ultrasmooth areas is drastically increased upon cooling to 140 K. This results in a reduction of $Rq$ from 2.9 (293 K) to 2.0 nm (140 K) as also highlighted in Figure S2 b and corroborated by the narrowing of the height distribution function (see Figure S3 b). Deposition at 140 K leads to well-defined interfaces and surfaces as evidenced by the low damping of Kiessig fringes in the XRR measurement (see Figure S3 c). In case of deposition at 293 K the Kiessig fringes are fully damped after a mere 9 oscillations. The reduced critical angle of total reflection is associated with a lower mean electron density within the film.[4] Substrate cooling not only affects the surface morphology but, in addition, also the crystallite structure of the thin films. The XRD patterns of the respective samples demonstrate an enhancement in crystallinity upon substrate cooling (see Figure S3 d). Cooling of the substrate to 140 K results in a well-defined crystallite growth perpendicular to the substrate plane, as evidenced by distinct (111) and (222) Bragg reflections. Other diffraction peaks are almost absent with only a residual (200) reflection being detected, despite the non-zero structure factors of the (200), (220) and (311) Bragg reflections for the face-centered cubic unit cell with monoatomic base. High resolution scans of the (111) Bragg reflection shown in Figure S3 e, confirm the high crystal quality for deposition at 140 K by the presence of distinct Laue oscillations.

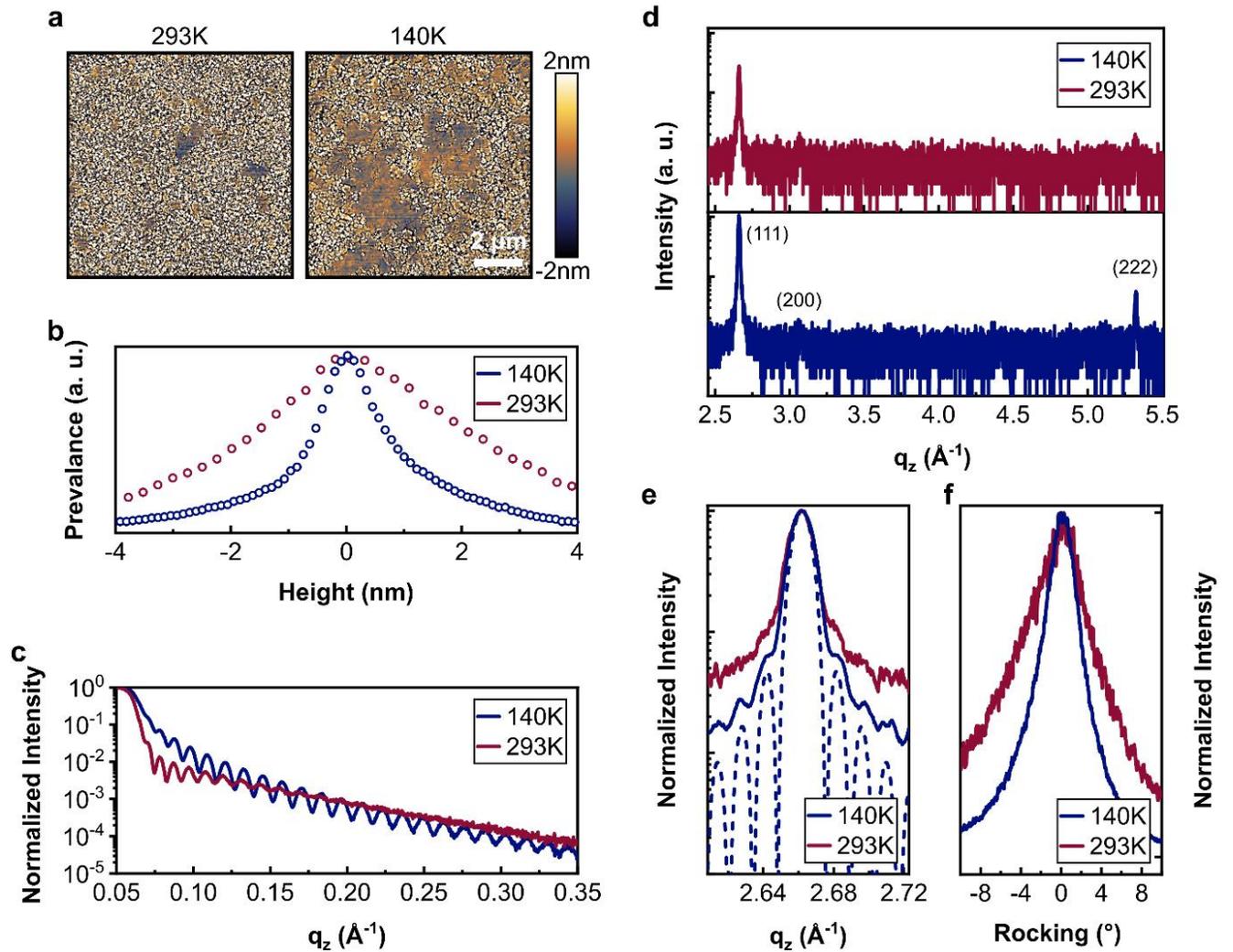

**Figure S3.** Characterization of the structural properties of Ag thin films deposited at substrate temperatures of 293 K and 140 K. a) Tapping mode AFM images of a representative 10 x 10 μm² area. The color bar is scaled from -2 to 2 nm normalized to the median height. *Rq* is reduced from 2.9 nm to 2.0 nm upon substrate cooling (see also Figure S2b). Prior to visualization a polynomial background correction was applied to the data. b) Height distribution functions deduced from the respective AFM images. The smooth growth at 140 K is evidenced by the narrower distribution. c) XRR data as an integral measure (22 x 22 mm²) of the surface and interface roughness. In case of thin films deposited at 293 K the Kiessig fringes are fully damped after only 9 oscillations. d) XRD patterns ($\theta$-$2\theta$ scan) of the thin films. The crystallinity of the thin film deposited at 140 K is enhanced as evidenced by the higher intensity of the (111) and (222) Bragg reflections. e) Semilogarithmic presentation of the normalized (111) Bragg reflections. The presence of distinct Laue oscillations (model function as dashed blue line) is indicative for a high crystallite quality and a low defect density in case of deposition at 140 K. f) Rocking curves on the (111) Bragg reflections indicating a strongly reduced crystallite mosaicity for 140 K. A moving average filter was applied to the (111) Bragg reflections and the rocking curves prior to data evaluation.

The simulated Laue model, as described in the main text, closely matches the experimentally observed periodicity and intensity distribution, under the assumption of a coherently diffracting domain size of 40 nm (170 lattice planes),. The enhanced crystallinity and crystallite quality for deposition at 140 K is accompanied by a reduced crystallite mosaicity, and hence, a more oriented growth along the substrate normal, which is confirmed by the reduced width of the rocking curves (see Figure S3 f).

It is imperative that elevated deposition rates and substrate cooling are employed, to facilitate the fabrication of high-quality Ag thin films. With a substrate temperature of 140 K and a deposition rate of 3 nm $s^{-1}$, a suitable set of parameters for the defined and controlled growth of Ag thin films has been established.

## SI. 2: Supporting Morphological Characterization of $Ag_{100-x}Au_x$ Thin Films

Thin film composition and, in particular, the homogeneous incorporation of Au into the $Ag_{100-x}Au_x$ thin films were confirmed by energy-dispersive X-ray (EDX) spectroscopy. The results of the EDX mapping are depicted in Figure S4.

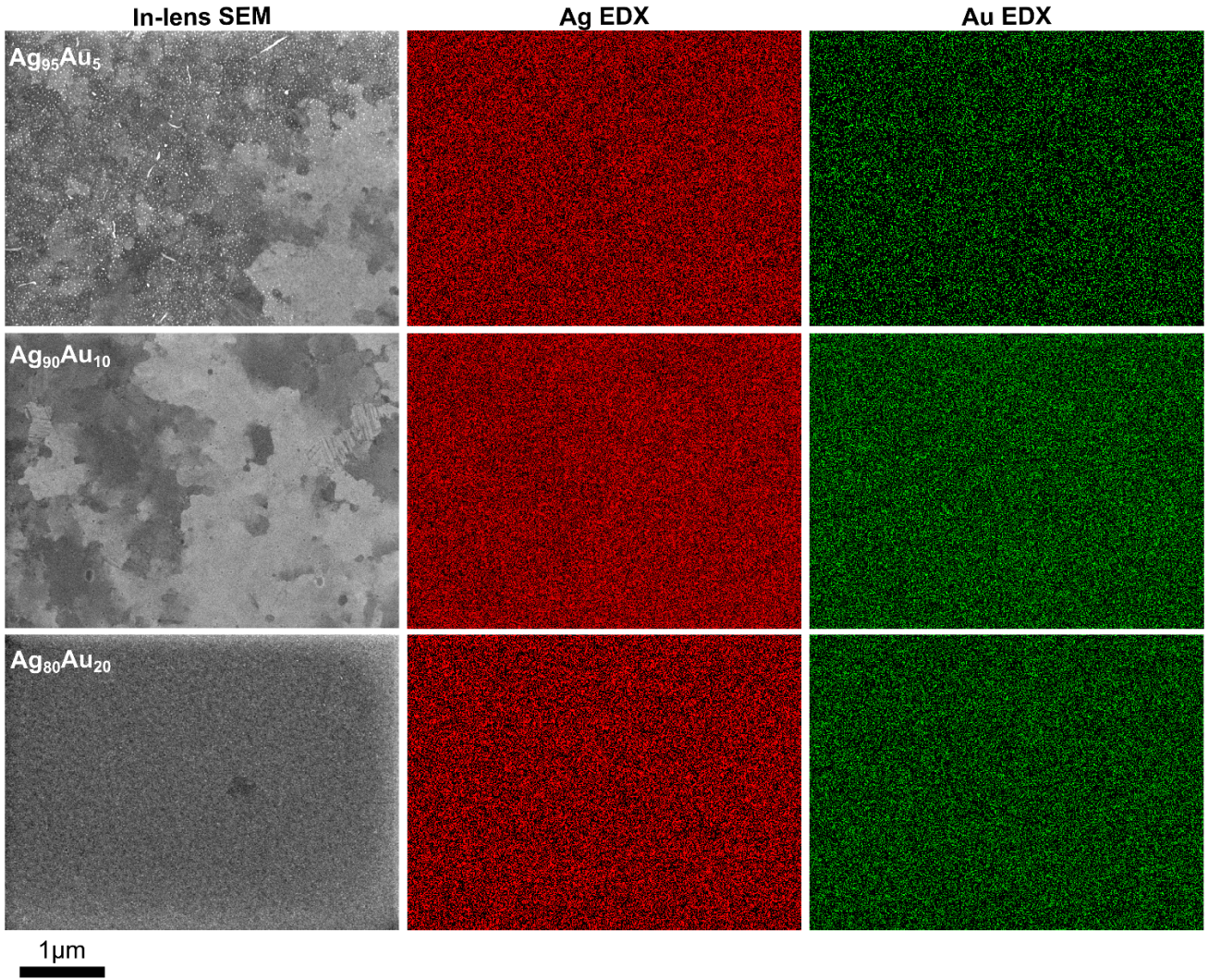

**Figure S4.** Energy-dispersive X-ray (EDX) mapping of $Ag_{95}Au_5$, $Ag_{90}Au_{10}$ and $Ag_{80}Au_{20}$ thin films (from top to bottom). Left column: In-lens SEM micrographs (acceleration voltage 5 kV, magnification 10,000). Middle column: Respective EDX mapping of Ag. Right column: Respective EDX mapping of Au. The maps were scaled to similar contrast values for better visibility. As evident from the maps Au is homogenously incorporated in the alloyed thin films. The composition of the thin films was determined by quantitative EDX spectroscopy on the same area.

A phenomenological function *g(r)* described by Gredig et al.[5] was employed to model the height-height correlation data extracted from the tapping mode AFM images displayed in Figure 1a:

$$g(r) = 2Rq^2 \left[1 - e^{-\left(\frac{r}{\xi}\right)^{2\alpha}}\right] \quad (SI1)$$

Here $Rq$ is the root mean square roughness of the film, $r$ is the distance between two points, $\xi$ is the correlation length and $\alpha$ is the Hurst parameter. The Hurst parameter defines the shape of the correlation function and is influenced by the grain growth mode. As the lateral resolution of the AFM scans limits the point distance to 20 nm finer details of the initial rise of the height-height correlation are not captured which makes the Hurst parameter prone to a large fit error. The height-height correlation functions of $Ag_{100-x}Au_x$ thin films, as extracted from the tapping mode AFM images displayed in Figure 1a, are presented in Figure S5, alongside the corresponding fit functions. The blue functions correspond to the entire area of 10 x 10 µm². The correlation length $\xi$ (see Table S1), is a measure for the average lateral surface grain size and smoothness. Across the 10 x 10 µm² surface area, all thin film samples exhibit a similar correlation length between 20 to 40 nm, except for the $Ag_{80}Au_{20}$ thin film, where the correlation length falls below the AFM resolution due to its fine-grained structure. The red functions represent a flat area of 1 x 1 µm².As expected for large crystallites, the correlation length increases significantly for the flat 1 x 1 µm² AFM scan section, coinciding with a substantial reduction in $Rq$. All extracted parameters are listed in Table S1.

**Table S1.** Surface properties of $Ag_{100-x}Au_x$ thin films derived from AFM measurements (Figure 1a). The correlation length $\xi$ and the scaling exponent $\alpha$ are determined by fitting the corresponding height-height correlation data. $Rq$ values were obtained by statistical analysis and are identical to those derived from height-height correlation analysis. For thin films that according to the visual appearance in Figure 1a exhibit two distinct growth modes (Ag, $Ag_{95}Au_5$ and Au) an additional analysis was performed on selected ultrasmooth regions (area: 1 x 1 µm², see markers in Figure 1a).

| Thin Film | 10 x 10 µm² | | | 1 x 1 µm² | | |
|---|---|---|---|---|---|---|
| | $Rq$ (nm) | $\xi$ (nm) | $\alpha$ | $Rq$ (nm) | $\xi$ (nm) | $\alpha$ |
| Ag | 2.0 | 32 ± 5 | 1.0 | 0.3 | 42 ± 5 | 0.6 |
| $Ag_{95}Au_5$ | 0.9 | 25 ± 5 | 0.7 | 0.4 | 31 ± 5 | 0.6 |
| $Ag_{90}Au_{10}$ | 0.5 | 26 ± 5 | 0.5 | / | / | / |
| $Ag_{80}Au_{20}$ | 0.5 | < 20 | / | / | / | / |
| Au | 0.6 | 27 ± 5 | 0.8 | 0.5 | 31 ± 5 | 0.6 |

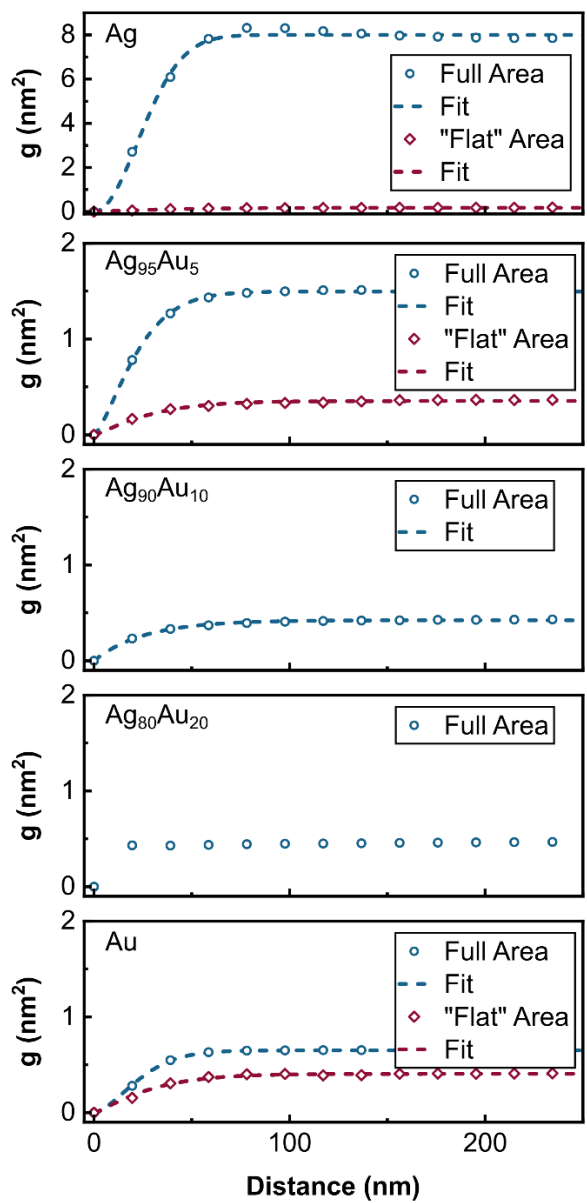

**Figure S5.** Height-height correlation functions of $Ag_{100-x}Au_x$ thin films extracted from the tapping mode AFM images shown in Figure 1a. The blue functions correspond to the full area of 10 x 10 μm². The red functions represent an area of 1 x 1 μm² with flat extended grains selected from the full area (see markers in Figure 1a). The respective fit functions (Equation SI1) are presented in dashed lines. The extracted parameters are listed in Table. S1.

To gain complementary information on the grain texture discussed on the basis of the AFM data presented in the main section (see Figure 1), in-lens SEM micrographs of the pure and alloyed thin films have been collected (see Figure S6).

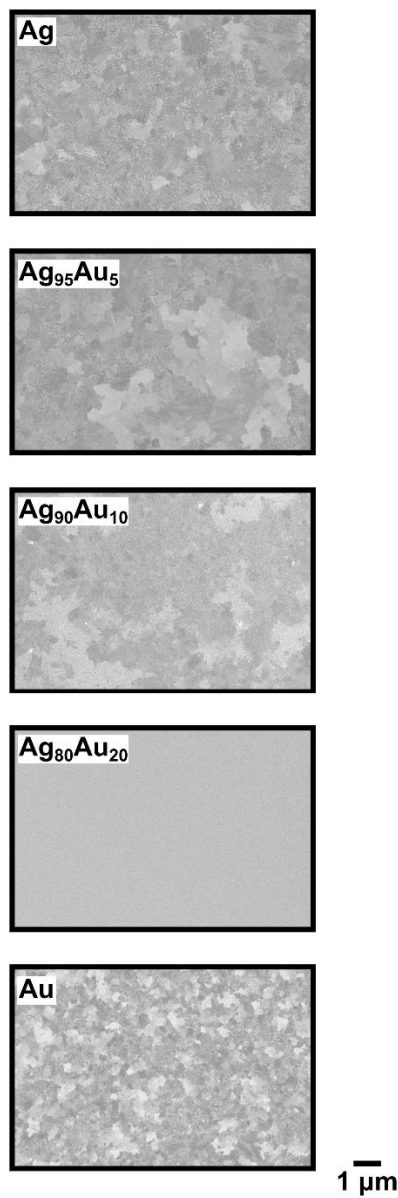

**Figure S6.** In-lens SEM micrographs (acceleration voltage 5 kV, magnification 10,000) of $Ag_{100-x}Au_x$ thin films. It becomes evident from the SEM images that alloying in the range of 5 to 10 at% results in predominantly extended grains of several micrometers in dimensions. A notable exception is the $Ag_{80}Au_{20}$ thin film, which exhibits grain sizes well below 1 µm, aligning with the height-height correlation and XRD data of the respective film.

AFM images are only representative for a selected sample area and the roughness at the substrate-film interface remains concealed. As a means, to obtain integral information on the surface morphology and interface roughness of our thin films, we have applied XRR measurements on the 22 x 22 mm$^2$ deposition area (see Figure S7).

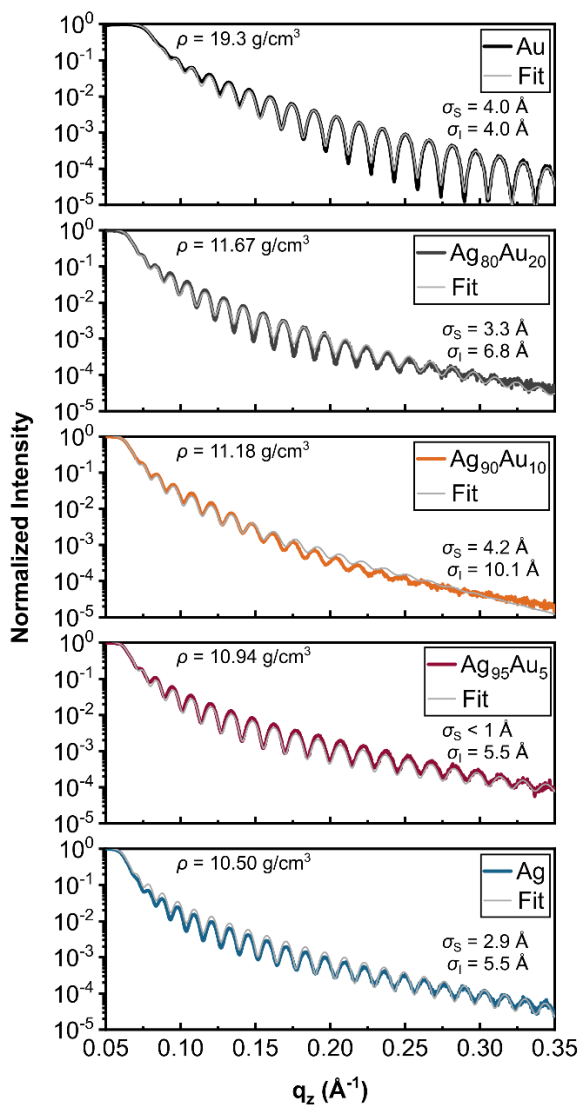

**Figure S7.** X-ray reflectivity data of $Ag_{100-x}Au_x$ thin films as an integral measure (22 x 22 mm$^2$) of the surface and interface roughness. The data were modelled by the GenX software package[6] under consideration of $SiO_2$ as substrate, a thin APTES layer and an Ag (Au) layer in air. The thin film density $\rho$, the surface roughness $\sigma_S$ and the interface roughness $\sigma_I$ were fitted as free parameters, as well as the metal thin film thickness (see Table S2). The $Ag_{95}Au_5$ thin film exhibits a remarkably low interface and surface roughness over the entire thin film area.

The presence of distinct Kiessig fringes with a low damping over the entire measurement range is indicative for low interface and surface roughness. Specifically, the Ag, Ag$_{95}$Au$_5$ and Au thin films feature a remarkably low Kiessig damping. Au is characterized by larger amplitudes of the Kiessig oscillations due to a higher electron density contrast at the interfaces. The XRR data were modelled by the GenX software package[6] to estimate the main structural parameters of each layer by the respective fits (displayed in grey in Figure S7) and, thereby, to quantify differences in interface and surface morphology. The fitted thin film densities are 10.50 g·cm$^{-3}$ (Ag), 10.94 g·cm$^{-3}$ (Ag$_{95}$Au$_5$), 11.18 g·cm$^{-3}$ (Ag$_{90}$Au$_{10}$), 11.67 g·cm$^{-3}$ (Ag$_{80}$Au$_{20}$) and 19.30 g·cm$^{-3}$ (Au). For comparison, the theoretical densities calculated based on the measured alloy compositions in at% are 10.93 g·cm$^{-3}$ (Ag$_{95}$Au$_5$), 11.37 g·cm$^{-3}$ (Ag$_{90}$Au$_{10}$), 12.26 g·cm$^{-3}$ (Ag$_{80}$Au$_{20}$). The comparison reveals that the Ag$_{90}$Au$_{10}$ and Ag$_{80}$Au$_{20}$ films exhibit slightly lower densities than theoretically expected. This deviation may be attributed to a combination of microstructural factors, including increased interface or grain boundary volume, porosity, and defects, also visible as holes in the AFM images. The Ag$_{95}$Au$_5$ thin film exhibits a remarkably low interface roughness of $\sigma_I$ = 5.5 Å and surface roughness of $\sigma_S$ < 1 Å across the entire film plane. This corresponds to a threefold reduction in surface roughness compared to the pure Ag film. In contrast to the localized AFM results, the Ag$_{90}$Au$_{10}$ and Ag$_{80}$Au$_{20}$ films show higher interface and surface roughness in the XRR analysis, highlighting the importance of complementary, large-area characterization. The thin film thicknesses determined via XRR are listed in Table S2.

**Table S2.** Thin film thicknesses of $Ag_{100-x}Au_x$ thin films obtained from XRR data (see Figure S7). Two different methods were applied to derive the thin film thickness. Either Parrat´s formalism was applied in GenX or the thin film thickness was determined from the distance of the Kiessig minima according to Spieß et al.[7].

| Thin Film | Parrat Fit (nm) | Spieß[7] (nm) |
|---|---|---|
| Ag | 46 ± 1 | 49 ± 3 |
| $Ag_{95}Au_5$ | 40 ± 1 | 43 ± 2 |
| $Ag_{90}Au_{10}$ | 40 ± 2 | 46 ± 3 |
| $Ag_{80}Au_{20}$ | 43 ± 1 | 47 ± 3 |
| Au | 38 ± 1 | 43 ± 2 |

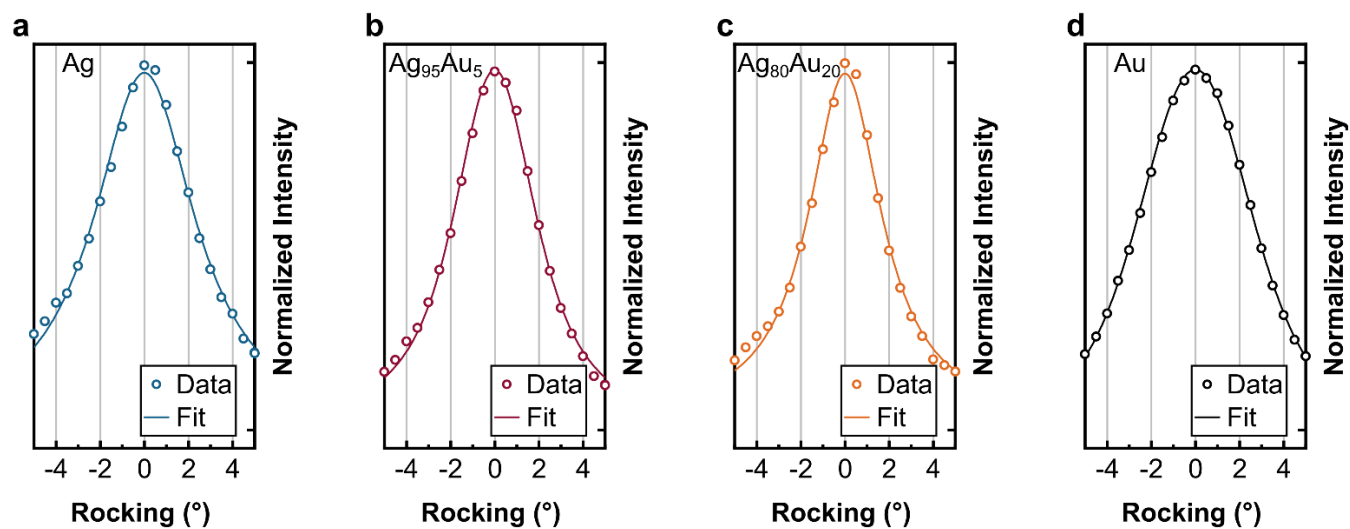

**Figure S8.** Rocking curves on the (111) Bragg reflection of $Ag_{100-x}Au_x$ thin films. Voigt functions were fitted to the data to retrieve the full width at half maximum (FWHM). a) Ag with a FWHM of 5.6°, b) $Ag_{95}Au_5$ with a FWHM of 4.5°, c) $Ag_{90}Au_{10}$ with a FWHM of 4.0°, d) Au with a FWHM of 6.1°. Au alloying of Ag thin films leads to a more oriented growth of crystallites with respect to the substrate plane.

## SI. 3: Supporting Optical Characterization of $Ag_{100-x}Au_x$ Thin Films

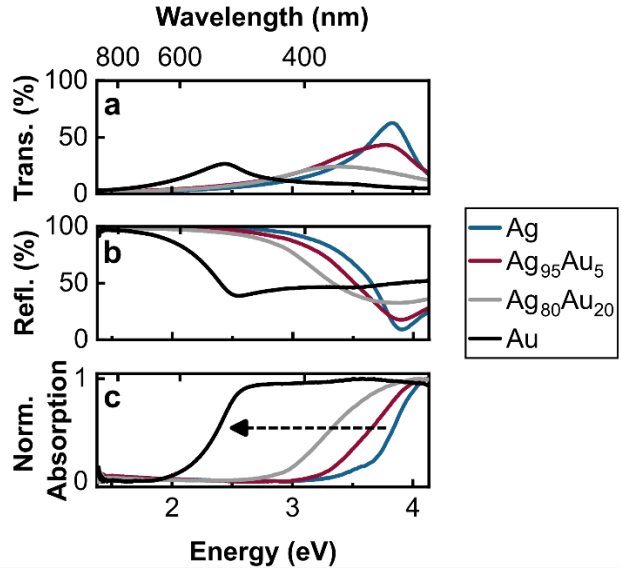

**Figure S9.** Optical properties of $Ag_{100-x}Au_x$ thin films. a) Measured transmission spectra. The transmission peaks are sensitive to the respective chemical composition of the film with peak maxima at 324 nm (Ag), 328 nm ($Ag_{95}Au_5$) and 369 nm ($Ag_{80}Au_{20}$). The Au thin film exhibits a dual transmission peak at 359 nm and 508 nm. b) Measured reflection spectra. c) Normalized absorption spectra calculated from the transmission and reflection spectra. The absorption edge of the interband transitions is sensitive to the Au content and shifts to lower energies with increasing Au content, thus increasingly blocking access to the visible spectral range.

**Table S3.** Real ($\varepsilon_1$) and imaginary component ($\varepsilon_2$) of the dielectric function for selected Ag thin films. The Ag and $Ag_{95}Au_5$ thin films reported in this work are compared to literature reported reference Ag systems at wavelengths of 500 nm and 600 nm.

|  | @500 nm | | @600 nm | | |
| --- | --- | --- | --- | --- | --- |
| **Thin Film** | $\varepsilon_1$ | $\varepsilon_2$ | $\varepsilon_1$ | $\varepsilon_2$ | **Fabrication Type** |
| *Ag (this work)* | *-9.37* | *0.32* | *-15.47* | *0.35* | as evaporated |
| *$Ag_{95}Au_5$ (this work)* | *-9.52* | *0.30* | *-15.79* | *0.46* | as evaporated |
| Johnson & Christy[8] | -9.80 | 0.31 | -16.07 | 0.44 | as evaporated |
| McPeak et al.[9] | -9.98 | 0.26 | -16.34 | 0.38 | template-stripped |
| Yang et al.[10] | -9.37 | 0.32 | -15.1 | 0.43 | template-stripped |

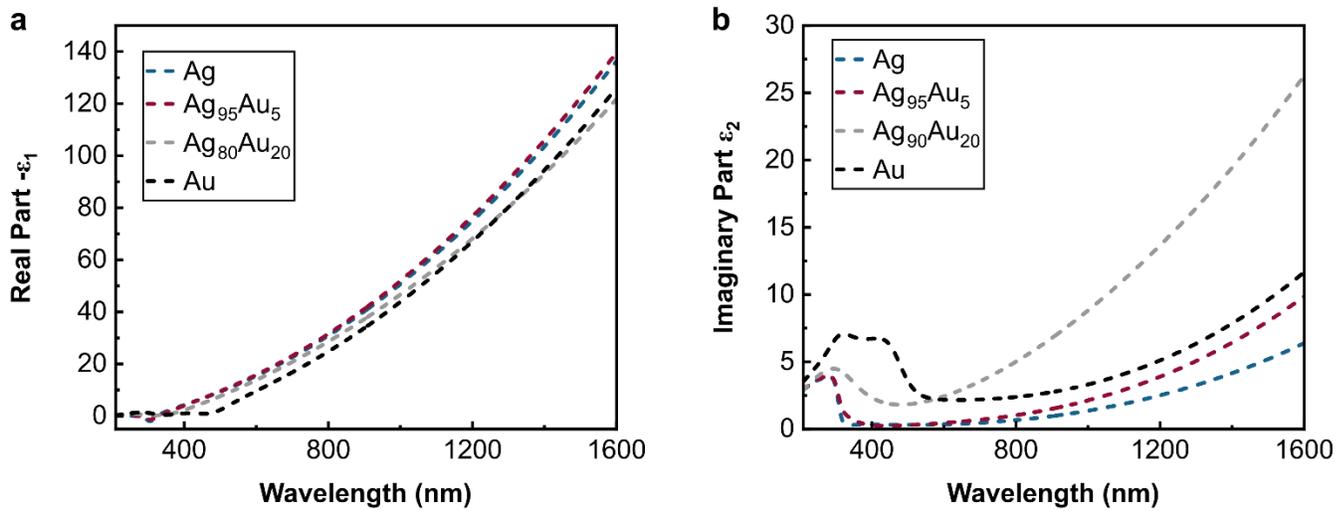

**Figure S10.** Dielectric functions of $Ag_{100-x}Au_x$ thin films measured by spectroscopic ellipsometry in a wavelength regime of 210 to 1600 nm. a) Negative real part. b) Imaginary part.

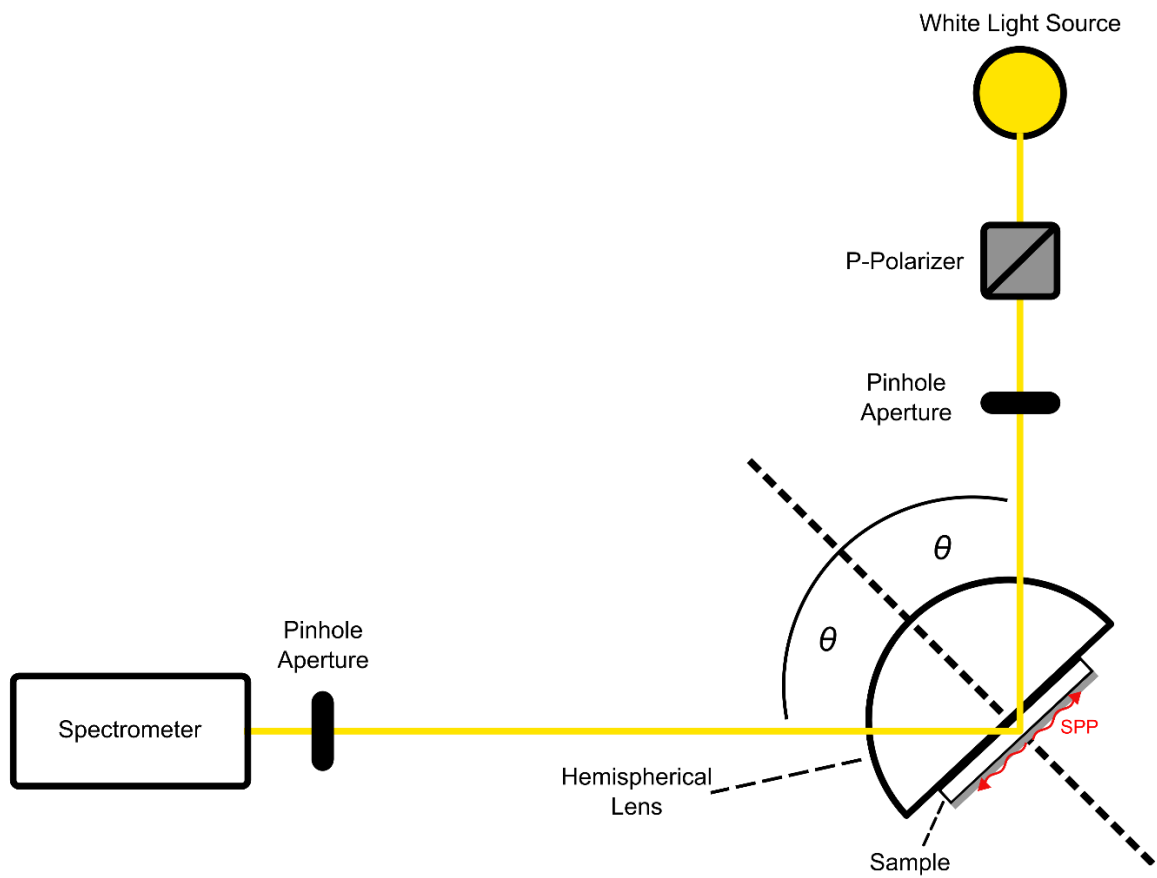

**Figure S11.** Schematic of the Kretschmann setup used for measurements of surface plasmon polariton (SPP) dispersions.

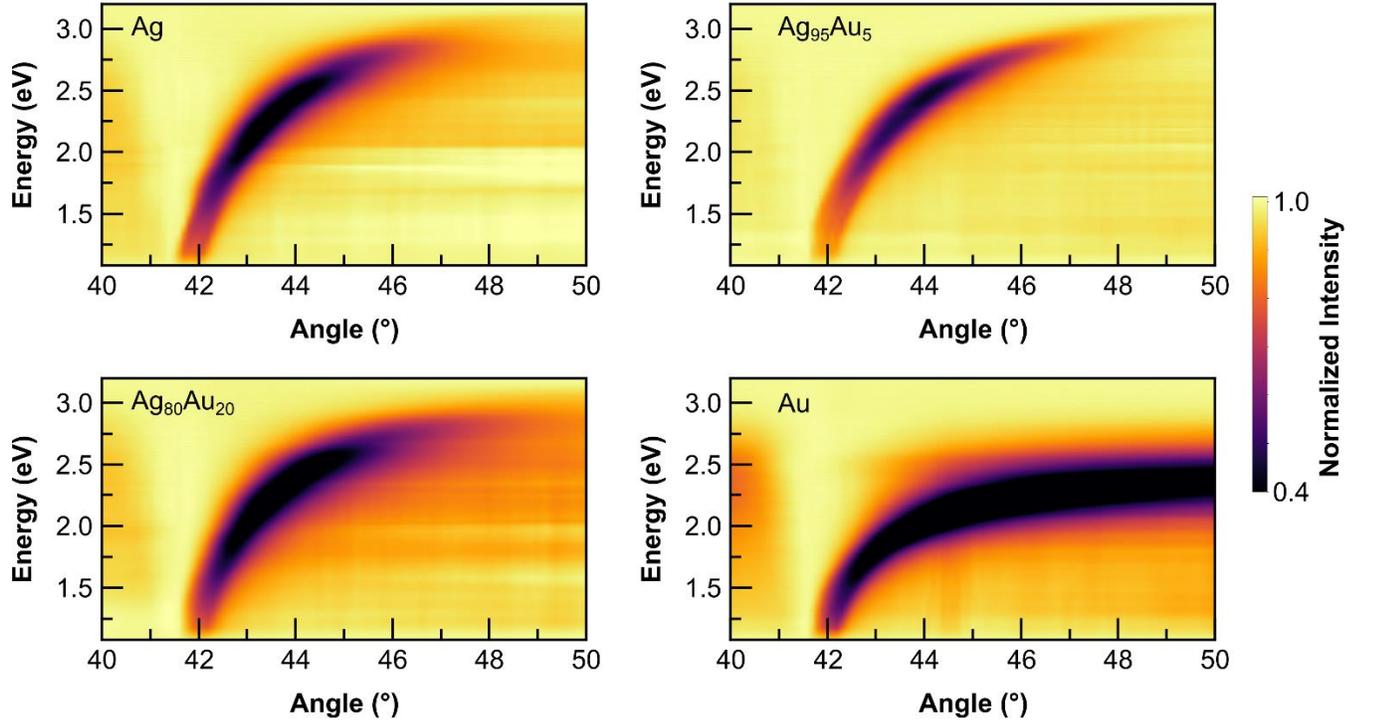

**Figure S12** Surface plasmon polariton (SPP) dispersions of the thin films measured in a Kretschmann configuration. The color bar refers to the reflected intensity as function of the energy and the polar angle with respect to the sample normal. A minimum in intensity is caused by SPP excitation.

The SPP dispersion relation between the wavevector $k$ along the direction of propagation and the angular frequency $\omega$ is given as follows:[11]

$$k = \frac{\omega}{c}\sqrt{\frac{\varepsilon_D \varepsilon_M}{\varepsilon_D + \varepsilon_M}} \qquad (SI2)$$

where $c$ is the speed of light in vacuum, $\varepsilon_D$ is the dielectric constant of air at room temperature and $\varepsilon_M$ is the complex dielectric function of the metal. When we estimate the real part of $\varepsilon_M$ with $\varepsilon'_M \approx \varepsilon_\infty - \omega_p^2/\omega^2$,[10] where $\varepsilon_\infty$ is the high frequency dielectric constant, due to interband transitions, and $\omega_p$ is the plasma frequency of the metal, we can derive the following relationship between the energy $E$ and the wavevector $k$:

$$E = \hbar \cdot c \cdot \sqrt{\frac{\left(\varepsilon_D\left(\frac{\omega_p}{c}\right)^2 + k^2(\varepsilon_D + \varepsilon_\infty)\right) - \sqrt{\left(\varepsilon_D\left(\frac{\omega_p}{c}\right)^2 + k^2(\varepsilon_D + \varepsilon_\infty)\right)^2 - 4\varepsilon_\infty \varepsilon_D k^2\left(\frac{\omega_p}{c}\right)^2}}{2\varepsilon_\infty \varepsilon_D}} \quad \text{(SI.3)}$$

Here $\hbar$ is the reduced Planck`s quantum of action. We have fitted Equation SI.3 to the minima positions in the reflection spectra as function of the wavevector $k$. The results are displayed in Figure S13.

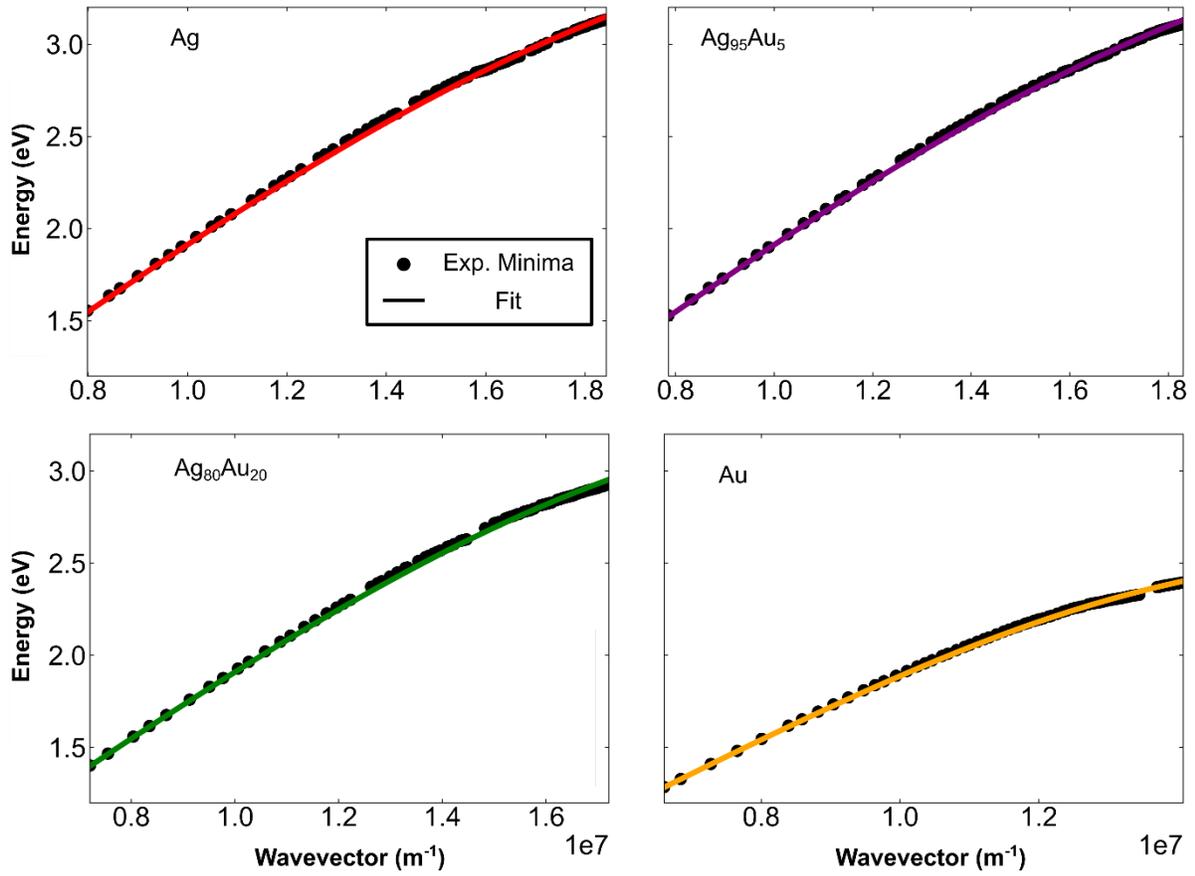

**Figure S13.** Fits of the intensity minima in the SPP dispersions according to Equation SI3.

## SI 4: Supporting Data on the Oxidative Stability Test

To assess chemical stability against oxidation, we have exposed Ag, $Ag_{95}Au_5$, $Ag_{80}Au_{20}$ and Au thin films to an oxidative environment by means of immersion in 0.2 % (w/w) aqueous $H_2O_2$ solution for 15 minutes. In-lens SEM and XRR measurements were taken before and after immersion. While SEM provides local surface information, XRR captures integral data across the entire 22 x 22 mm$^2$ sample area. The SEM micrographs (left column) and XRR data (right column) before and after the degradation test are presented in Figure S14.

As demonstrated by the SEM micrographs (Figure S14a), the pure Ag film is highly susceptible to oxidation, evident from large areas of film removal and significant changes in grain texture. This material loss is confirmed by XRR (Figure S14b) through a shift in the edge of total reflection to a smaller momentum transfer. The shift in the edge of total reflection could also be attributed to the incorporation of oxygen or the creation of vacancies in the film. The stronger damping of Kiessig fringes further indicates increased surface roughness from degradation.

Impressively so, the presence of Au in $Ag_{95}Au_5$ thin films significantly enhances the stability against oxidation, as proven by the absence of material removal as well as grain texture modification in the SEM micrographs (Figure S14c). Furthermore, the edge of total reflection in the XRR measurements (Figure S14d) remains fixed, indicating that no significant amounts of oxygen are incorporated, and no substantial fractions are removed from the bulk film. The decrease in the amplitude and frequency of Kiessig oscillations suggests that the surface of the thin film only slightly reacts in the solution.

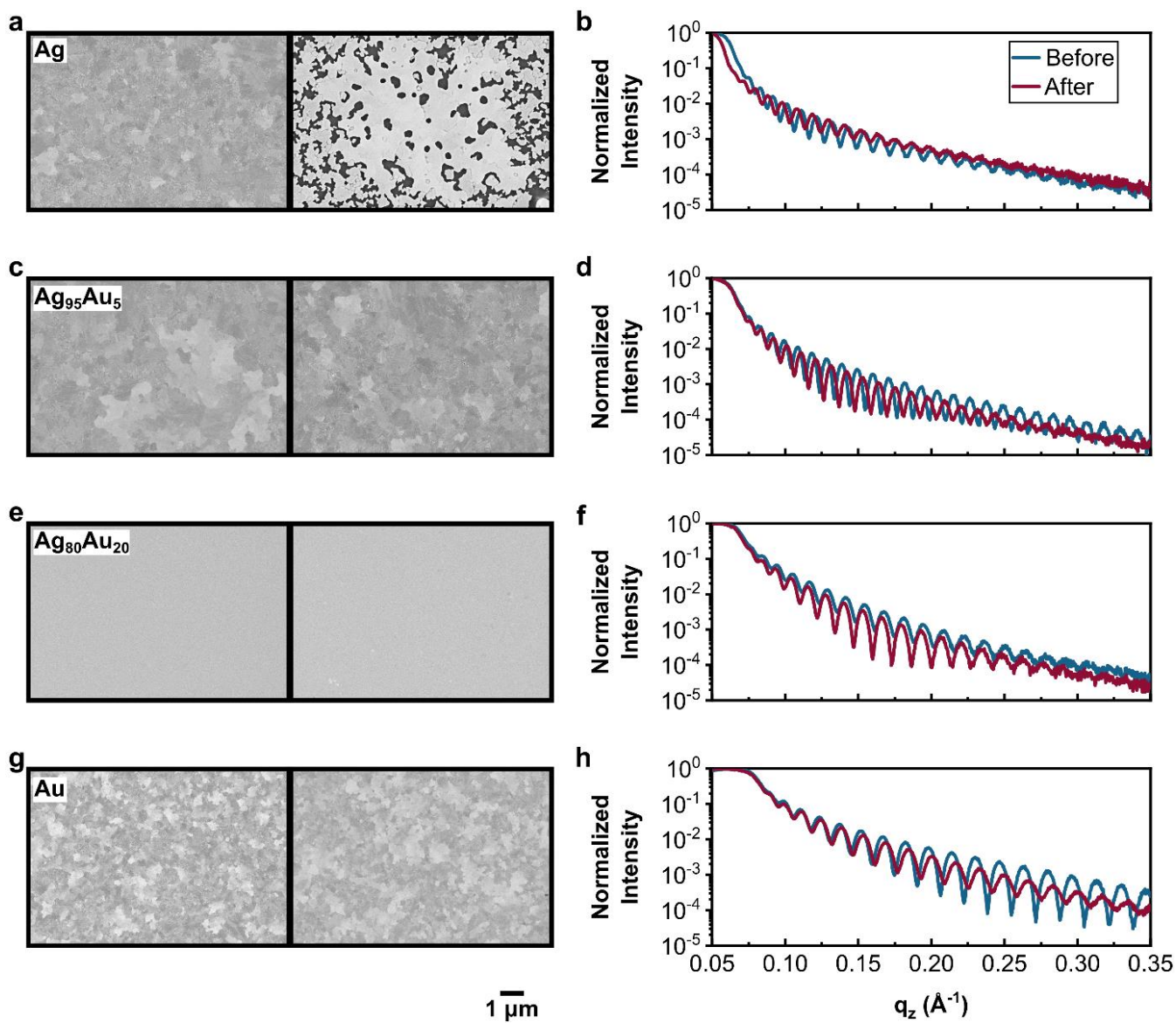

**Figure S14.** Oxidative stability of $Ag_{100-x}Au_x$ thin films. The thin films were immersed (15 minutes) into 0.2 % (w/w) aqueous $H_2O_2$ solution. In the left column, representative in-lens SEM micrographs (acceleration voltage 5 kV) are shown before (left) and after immersion (right). The right column shows XRR curves before and after immersion. Ag (a, b), $Ag_{95}Au_5$ (c, d), $Ag_{80}Au_{20}$ (e, f) and Au (g, h) thin film properties are shown from top to bottom.

Interestingly the $Ag_{80}Au_{20}$ film Figure S14e and f) displays a similar level of chemical stability despite its different, fine-grained morphology and lack of (111) texture. This implies, that the enhanced oxidation resistance results primarily from Au incorporation, not just structural quality.

A comparison of $Ag_{95}Au_5$ (Figure S14c and d) with pure Au (Figure S14g and h), highlights the chemical stability of our alloyed thin films, as even for the pure Au thin film a slight reaction of the surface in solution is observed, which is in turn visible from an increased damping of the Kiessig oscillations. The reduction in overall thin film thickness during immersion remains within 1 nm for all concentrations studied.